\documentstyle[amssymb,preprint,aps,floats]{revtex}
\tightenlines
\begin{document}
\title{A Simple Treatment of Metal-Insulator Transition: Effects of Degeneracy,
Temperature and Applied Magnetic Field}
\author{Andrzej Klejnberg and Jozef Spa\l ek}
\address{Marian Smoluchowski Institute of Physics, Jagiellonian University,\\
ulica Reymonta 4, 30-059 Krak\'{o}w, Poland}
\maketitle

\begin{abstract}
A simple slave-boson representation combined with the Hartree-Fock
approximation for the Hund's rule coupling is introduced for a doubly
degenerate narrow band, which bears a direct relation to that introduced
previously in the nondegenerate case. Namely, one keeps the fermion
representation of the spin operator to recover properly the energy of
fermionic quasiparticles in the presence of an applied magnetic field. A
simple two-parameter mean-field analysis of the metamagnetism is provided,
with the emphasis on the role of the Hund's rule coupling. We also analyse
the appearance of the spin-split effective masses in the applied field and
for nonhalf-filled-band situation. The Mott-Hubbard boundary is determined
at nonzero temperature ($T>0$); it shifts towards lower interactions with
increasing $T$ and the field signalling the precursory localization effects,
explicitly exhibited in the behavior of the magnetic susceptibility
calculated in the Appendix. We also formulate a more general two-parameter
rotationally invariant approach for an arbitrary degeneracy $d$ of
equivalent orbitals and show that the Mott-Hubbard transition at zero
temperature and at any integer filling $n>1$ is always discontinuous. A
brief overview of experimental situation is also made.
\end{abstract}

\section{Introduction}

In recent years one observes a renewed interest in the metal-insulator
transition of the Mott-Hubbard type, with a particular emphasis on the
effect of orbital degeneracy and associated with it Hund's rule coupling,
both appearing as additional factors \cite{1}. This interest is induced by
the new works on perovskites \cite{2}, NiS$_{2-x}$Se$_x$ \cite{3}, as well
as on the canonical system V$_{2-y}$O$_3$ \cite{4}. In most of these
compounds one observes an anomalous metallicity even in the paramagnetic
state \cite{3,4}. Among them is the metamagnetism and a relatively strong
applied magnetic field dependence of thermodynamic properties \cite{5}, in
addition to the field induced metal-insulator transition \cite{6}. These
phenomena of localization occur in compounds of various crystal structure
and even in the same magnetic phase, as in the case of NiS$_{2-x}$Se$_x$.
Therefore, it seems that the underlying microscopic mechanism is rather
general, neither strongly dependent on a particular crystallographic
structure, nor on the form of the density of states near the Fermi energy
and the type of magnetism. Additionally, electronic properties of the system
NiS$_{2-x}$Se$_x$ are well described by considering a half-filled $e_g$ band 
\cite{7} composed of $3d$ states (due to Ni$^{2+}$ ions), which hybridize
with completely filled $p$ states due to S$^{2-}$ or Se$^{2-}$ (the last
states are believed to play only a passive role near MIT).

The purpose of this paper is to carry out a simplified analysis for a system
composed of orbitally degenerate but otherwise equivalent band states that
nonetheless addresses some of the principal phenomena in an applied magnetic
field and at nonzero temperature. Thus, one can examine the system behavior
as a function of experimentally controllable parameters.

The analysis of Mott-Hubbard localization in the nondegenerate band case was
based on the Hubbard model and provides a continuous transition at $T=0$ 
\cite{8} and a discontinuous transition at $T>0$ \cite{9}. These results
were subsequently confirmed in the limit of infinite dimensions \cite{10}.
This means that this quantum phase transition possesses an upper critical
dimension and the Gutzwiller approach represents a correct mean-field
theory. Also, close to the localization, depending on the band filling, one
observes metamagnetism \cite{6} or metamagnetic behavior \cite{7,11},
spin-split effective masses \cite{11,12}, and a transition from an
antiferromagnetic semimetal to an antiferromagnetic insulator \cite{4}. The
related Gutzwiller approximation scheme for a doubly degenerate Hubbard
model provides a discontinuous transition already at $T=0$ \cite{1}, a
change induced by the Hund's rule coupling. Obviously, the Hund's rule
should make the localization easier (i.e. diminish the critical value $U_c$
of the intraatomic interaction), since it favors energetically the high-spin
atomic state against any normal Fermi-liquid state. Hence, we can separate
the effects associated with the Coulomb interaction from those due to the
Hund's rule coupling.

The structure of this paper is as follows. In the next section we introduce
a slave-boson approach combined with the Hartree-Fock approximation for the
Hund's-rule term, which reproduces in a simple manner the similar results of
the full slave-boson analysis by Hasegawa \cite{1} in the paramagnetic state
for zero field and temperature. Additionally, this representation provides a
correct expression of the energy a fermionic quasiparticles in the applied
magnetic field. Within this scheme we determine the magnetic susceptibility,
metamagnetic properties, spin-dependent effective masses, as well as
determine the Mott-Hubbard boundary at nonzero temperature, and its shift
towards lower temperatures in the applied field. The similarities and
differences with the nondegenerate-band case are stressed. In the second
part we propose a simple rotationally invariant formulation of the problem
for an arbitrary orbital degeneracy of electron states and compare it with
the Gutzwiller approach. Within this approach, a difference between the
Hund's rule and intersite magnetic coupling appears naturally, although the
last point is not discussed in detail in the present paper. The correlated
metal-paramagnetic insulator transition takes place at any integer filling $%
n $ and is discontinuous for $n>1.$
\newpage
\section{The para- and meta-magnetic states of doubly degenerate Hubbard
model.}

\subsection{The slave-boson method combined with the Hartree-Fock
approximation for the exchange term.}

Doubly degenerate Hubbard model in the situation with equivalent orbitals is
described by Hamiltonian:

\begin{equation}
{\cal H}=\sum_{ijl\sigma }{}^{\prime }t_{ij}a_{il\sigma }^{\dagger
}a_{jl\sigma }+U\sum_{il}n_{il\uparrow }n_{il\downarrow }+\left( U^{^{\prime
}}-\frac 12J\right) \sum_in_{i1}n_{i2}-2J\sum_i{\bf S}_{i1}\cdot {\bf S}%
_{i2}-2h\sum_{il}S_{il}^z\ ,  \label{hub}
\end{equation}
where $a_{il\sigma }$ is the annihilation operator of electron from lattice
site $i$ on orbital $l(=1,2)$ and with spin $\sigma (=\uparrow ,\downarrow
), $ $n_{il\sigma }$ is the particle number and ${\bf S}%
_{il}=(S_{il}^{+},S_{il}^{-},S_{il}^z)$ is the spin operator, $t_{ij}$ is
the hopping integral, assumed for simplicity the same for the two orbitals
(we choose also $t_{ii}=0$). Parameters $U,$ $U^{^{\prime }}$ and $J,$ are
respectively, the intraband Coulomb, the interband Coulomb and the
Hund's-rule-exchange integrals. For $e_g$ band we have: $U^{^{\prime
}}=U-2J. $ Finally, in the last term $h=\frac 12g\mu _BH_a,$ where $H_a$ is
the applied magnetic field. Because of the vanishing orbital moment in the
first order, the magnetic field affects the spin degrees only (we neglect
also the Landau quantization effects for low fields).

We make the mean-field approximation for the exchange term. Likewise, we
decouple in the same manner the interorbital term, since $U^{^{\prime }}-%
\frac 12J=U-\frac 52J.$ Therefore, we have

\begin{eqnarray}
n_{i1}n_{i2} &=&n_{i1}\langle n_{i2}\rangle +\langle n_{i1}\rangle
n_{i2}-\langle n_{i1}\rangle \langle n_{i2}\rangle \ ,  \nonumber \\
{\bf S}_{i1}\cdot {\bf S}_{i2} &=&S_{i1}^z\langle S_{i2}^z\rangle +\langle
S_{i1}^z\rangle S_{i2}^z-\langle S_{i1}^z\rangle \langle S_{i2}^z\rangle \ ,
\label{hrtrb}
\end{eqnarray}
where $\langle A\rangle =Tr(Ae^{-\beta {\cal H}_{H-F}})$ is the average
value of operator $A$ with Hamiltonian ${\cal H}_{H-F}$:

\begin{eqnarray}
{\cal H}_{H-F} &=&\sum_{ijl\sigma }{}^{\prime }t_{ij}a_{il\sigma }^{\dagger
}a_{jl\sigma }+U\sum_{il}n_{il\uparrow }n_{il\downarrow }+K\frac 12%
n\sum_{il}n_{il}-Jm\frac 14\sum_{il\sigma }\sigma n_{il\sigma
}-h\sum_{il\sigma }\sigma n_{il\sigma }+  \nonumber \\
&&-\frac 14KNn^2+\frac 18JNm^2-\mu N_e\ .  \label{hubhrtr}
\end{eqnarray}
In this equation we have added the chemical potential part $(-\mu N_e),$
with $N_e$ being the number of electrons in the system. The $n$ and $m$ are
defined by equations:

\begin{eqnarray}
\langle n_{il}\rangle &\equiv &\frac 12n\text{ ,}  \label{n} \\
2\langle S_{il}^z\rangle &\equiv &\frac 12m\text{ ,}  \label{Sz}
\end{eqnarray}
and $N$ is the number of sites, $K=U^{^{\prime }}-\frac 12J.$ The mean
values of $n_{il}$ and $S_{il}^z$ are independent of site and orbital
indexes because of translational symmetry and the equivalence of the
orbitals for half- or nearly-half-filled band configurations (in this part
we consider only the limit of $n$ equal to or close to two).

We employ the slave boson representation of Kotliar and Ruckenstein \cite{13}
in the form:

\begin{mathletters}
\begin{eqnarray}
|0,il\rangle &\equiv &e_{il}^{\dagger }|v\rangle \ ,  \label{def_e} \\
a_{il\sigma }^{\dagger }|0,il\rangle &\equiv &|\sigma ,il\rangle \equiv
f_{il\sigma }^{\dagger }p_{il\sigma }^{\dagger }|v\rangle \ ,  \label{def_p}
\\
\sigma a_{il\sigma }^{\dagger }a_{il\bar{\sigma}}^{\dagger }|0,il\rangle
&\equiv &|2,il\rangle \equiv \sigma f_{il\sigma }^{\dagger }f_{il\overline{%
\sigma }}^{\dagger }d_{il}^{\dagger }|v\rangle \ ,  \label{def_d}
\end{eqnarray}
where we introduced new boson-fermion vacuum $|v\rangle $ instead of the
Fock-space vacuum $\{|0,il\rangle \}$. So, we have added only the orbital
index to the original formulation. The operators $e_{il}^{\dagger }$, $%
p_{il\sigma }^{\dagger }$ and $d_{il}^{\dagger }$ are boson operators and $%
f_{il\sigma }^{\dagger }$ is fermion operator. This representation must be
supplemented by the constraints

\end{mathletters}
\begin{equation}
Q_{il\sigma }\equiv f_{il\sigma }^{\dagger }f_{il\sigma }-p_{il\sigma
}^{\dagger }p_{il\sigma }-d_{il}^{\dagger }d_{il}\ =0  \label{con_1}
\end{equation}
and

\begin{equation}
P_{il}\equiv e_{il}^{\dagger }e_{il}+\sum_\sigma p_{il\sigma }^{\dagger
}p_{il\sigma }+d_{il}^{\dagger }d_{il}-1=0\   \label{con_2}
\end{equation}

Relation between original $\left( a_{il\sigma }\right) $ and new $\left(
f_{il\sigma }\right) $ fermion operators is:

\begin{equation}
a_{il\sigma }={\bf (}e_{il}^{\dagger }p_{il\sigma }+d_{il}p_{il\overline{%
\sigma }}^{\dagger })f_{il\sigma }=z_{il\sigma }f_{il\sigma }\ ,  \label{a_f}
\end{equation}

The constraints given by eqs. (\ref{con_1}) and (\ref{con_2}) can be
enforced by Lagrange multipliers $\lambda _{il}^{\left( 1\right) }$ and $%
\lambda _{il\sigma }^{\left( 2\right) }.$ We obtain following Hamiltonian:

\begin{eqnarray}
{\cal H}_{tot} &\equiv &{\cal H}_{H-F}+{\cal H}_{con}=\sum_{ijl\sigma
}{}^{\prime }t_{ij}z_{il\sigma }^{\dagger }z_{jl\sigma }f_{il\sigma
}^{\dagger }f_{jl\sigma }+U\sum_{il}d_{il}^{\dagger }d_{il}+  \nonumber \\
&&+\sum_{il\sigma }\left[ (\frac 12Kn-\mu )-\sigma (h+\frac 14Jm)\right]
f_{il\sigma }^{\dagger }f_{il\sigma }+{\cal H}_{con}+C\text{ ,}  \label{Htot}
\end{eqnarray}
where

\begin{equation}
C\equiv -\frac 14KNn^2+\frac 18JNm^2\ \text{,}  \label{Const}
\end{equation}
and

\begin{equation}
{\cal H}_{con}\equiv \sum_{il}\lambda _{il}^{\left( 1\right)
}P_{il}+\sum_{il\sigma }\lambda _{il\sigma }^{\left( 2\right) }Q_{il\sigma
}\ .  \label{Hwiez}
\end{equation}

Note that using the fermion representation for the spin operator $S_{il}^z=%
\frac 12(f_{il\uparrow }^{\dagger }f_{il\uparrow }-f_{il\downarrow
}^{\dagger }f_{il\downarrow })$, we obtain correctly the Zeeman term and
also, the Hund's rule coupling in the form of a molecular field.
Equivalently, one could utilize the Bose representation $S_{il}^z=\frac 12%
\left( p_{il\uparrow }^{\dagger }p_{il\uparrow }-p_{il\downarrow }^{\dagger
}p_{il\downarrow }\right) ,$ but then only the z-component is well defined
and the result in the saddle-point approximation does not provide the
quasiparticle energies with the proper Zeeman term (see the discussions of
this point in Ref.6b, where the full spin rotation invariant form of the SB
representation is invoked).

The factor $z_{il\sigma }$ is not unique, but when we make the choice \cite
{13}:

\begin{equation}
z_{il\sigma }\longrightarrow (1-d_{il}^{\dagger }d_{il}-p_{il\sigma
}^{\dagger }p_{il\sigma })^{-1/2}z_{il\sigma }(1-e_{il}^{\dagger
}e_{il}-p_{il\overline{\sigma }}^{\dagger }p_{il\overline{\sigma }})^{-1/2},
\label{zz}
\end{equation}
we obtain correct result in the saddle-point solution.

The partition function is given by:

\begin{equation}
Z=\int D[e,e^{\dagger }]D[p,p^{\dagger }]D[d,d^{\dagger }]D[f,f^{\dagger
}]\prod_{il\sigma }d\lambda _{il}^{\left( 1\right) }d\lambda _{il\sigma
}^{\left( 2\right) }\exp \left[ -\int_0^\beta {\cal L}(\tau )d\tau \right] \
,  \label{Z}
\end{equation}
where $\beta =1/k_BT$ is the inverse temperature,

\begin{eqnarray}
{\cal L}(\tau ) &=&\sum_{ijl\sigma }f_{il\sigma }^{\dagger }\left[ \left(
\partial _\tau -\widetilde{\mu }-\sigma (h+\frac 14Jm)+\lambda _{il\sigma
}^{\left( 2\right) }\right) \delta _{ij}+t_{ij}z_{il\sigma }^{\dagger
}z_{jl\sigma }\right] f_{jl\sigma }+  \nonumber \\
&&+\sum_{il}e_{il}^{\dagger }\left( \partial _\tau +\lambda _{il}^{\left(
1\right) }\right) e_{il}+\sum_{il\sigma }p_{il\sigma }^{\dagger }\left(
\partial _\tau +\lambda _{il}^{\left( 1\right) }-\lambda _{il\sigma
}^{\left( 2\right) }\right) p_{il\sigma }+  \nonumber \\
&&+\sum_{il}d_{il}^{\dagger }\left( \partial _\tau +\lambda _{il}^{\left(
1\right) }-\sum_\sigma \lambda _{il\sigma }^{\left( 2\right) }+U\right)
d_{il}-\sum_{il}\lambda _{il}^{\left( 1\right) }+C\ ,  \label{L}
\end{eqnarray}
is the system Lagrangian, and

\begin{equation}
\widetilde{\mu }=\mu -\frac 12Kn\ .  \label{tii}
\end{equation}
is the effective chemical potential.

Next, we make the saddle-point approximation, i.e. assume, that Bose fields
and Lagrange multipliers do not depend on time $\tau ,$ number of site $i$
and orbital index $l,$ so $e_{il}(\tau )\rightarrow e,$ $p_{il\sigma }(\tau
)\rightarrow p_\sigma ,$ $d_{il}(\tau )\rightarrow d$, $\lambda _{il\sigma
}^{\left( 2\right) }\rightarrow \lambda _\sigma ^{\left( 2\right) }$ and $%
\lambda _{il}^{\left( 1\right) }\rightarrow \lambda .$ To obtain
saddle-point solution we must minimize free energy with respect boson fields
and Lagrange multipliers. From partition function (\ref{Z}) we obtain
following expression for the free energy:

\begin{eqnarray}
F &=&-k_BT\sum_{{\bf k}l\sigma }\ln \left[ 1+\exp \left( \frac{\widetilde{%
\mu }-E_{{\bf k}\sigma }}{k_BT}\right) \right] +\frac 18NJm^2+2NUd^2+ 
\nonumber \\
&&-2N\sum_\sigma \lambda _\sigma ^{\left( 2\right) }\left( p_\sigma
^2+d^2\right) +2N\lambda ^{\left( 1\right) }\left( e^2+\sum_\sigma p_\sigma
^2+d^2-1\right) +  \nonumber \\
&&-\frac 14NKn^2+\mu Nn\text{ ,}  \label{F}
\end{eqnarray}
Fermi quasiparticle energy $E_{{\bf k}\sigma }$ is given by:

\begin{equation}
E_{{\bf k}\sigma }=q_\sigma \epsilon _{{\bf k}}-\sigma (h+\frac 14%
Jm)+\lambda _\sigma ^{\left( 2\right) },  \label{Ek}
\end{equation}
where $\epsilon _{{\bf k}}$ is bare band energy, $q_\sigma \equiv z_\sigma
^2 $ and

\begin{equation}
z_\sigma ^2=\frac{(ep_\sigma +dp_{\overline{\sigma }})^2}{(1-d^2-p_\sigma
^2)(1-e^2-p_{\overline{\sigma }}^2)}.  \label{z}
\end{equation}

Thus, one encounters here two molecular fields: one coming from the Hund's
rule and the other $\left( \lambda _{\uparrow }^{\left( 2\right) }-\lambda
_{\downarrow }^{\left( 2\right) }\right) $ coming from the electronic
correlations.

Differentiation of the free energy with respect Lagrange multipliers
provides the mean field version of the constraints:

\begin{eqnarray}
\frac{\partial F}{\partial \lambda _\sigma ^{\left( 2\right) }} &=&0\ \
\Rightarrow \ \ p_\sigma ^2+d^2=\frac 1N\sum_{{\bf k}}n_{{\bf k}\sigma }=%
\frac 12n_\sigma \text{ ,}  \label{df2} \\
\frac{\partial F}{\partial \lambda ^{\left( 1\right) }} &=&0\ \ \Rightarrow
\ \ 1-e^2-\sum_\sigma p_\sigma ^2-d^2=0\text{ ,}  \label{df1}
\end{eqnarray}
where

\begin{equation}
n_{{\bf k}\sigma }=\frac 1{1+\exp \left[ \beta \left( E_{{\bf k}\sigma }-%
\widetilde{\mu }\right) \right] }  \label{nk}
\end{equation}
is Fermi-Dirac function. From eqs. (\ref{n}) and (\ref{Z}) we obtain that $%
\langle n_{il\sigma }\rangle =\frac 12n_\sigma =(n+\sigma m)/4.$ Instead of
computing the derivatives with respect to the remaining fields we, put $e$
and $p_\sigma $ expressed via $d,$ $n$ and $m$ into free energy (\ref{F})$.$
Next, we introduce new variables $\beta $ and $\beta _0,$ such that $\lambda
_\sigma ^{\left( 2\right) }=\beta _0+\sigma \beta .$ We obtain the following
free energy

\begin{equation}
\frac FN=-2k_BT\frac 1N\sum_{{\bf k}\sigma }\ln \left[ 1+\exp \left( \frac{%
\overline{\mu }-E_{{\bf k}\sigma }}{k_BT}\right) \right] +\frac 18%
Jm^2+2Ud^2-\beta m+\frac 14Kn^2+\overline{\mu }n\text{ .}  \label{F/2N}
\end{equation}
Quasiparticle energy is given by:

\begin{equation}
E_{{\bf k}\sigma }=q_\sigma \epsilon _{{\bf k}}-\sigma (h+\frac 14Jm-\beta )%
\text{ ,}  \label{Ek_2}
\end{equation}
and effective chemical potential by:

\begin{equation}
\overline{\mu }=\widetilde{\mu }-\beta _0=\mu -\frac 12Kn-\beta _0\text{ .}
\label{mi_2}
\end{equation}

From (\ref{z}) we obtain $q_\sigma $ in a form

\begin{equation}
q_\sigma =4\frac{\left( \sqrt{1-\frac 12n+d^2}\sqrt{n+\sigma m-4d^2}+d\sqrt{%
n-\sigma m-4d^2}\right) ^2}{\left( 4-n-\sigma m\right) \left( n+\sigma
m\right) }  \label{qs}
\end{equation}

This factor is the mass enhancement and in general case $\left( n\neq
2\right) $ leads to the spin-dependent effective masses.

One should justify the approximations made. First, for $e_g$ band $J\simeq
0.2U$ , and therefore $U^{^{\prime }}=U-2.5J\simeq 0.5U.$ Hence one can say
that making the Hartree-Fock approximation for the interorbital-interaction
part is qualitatively correct even for $U\sim W.$ Below we make this
argument more quantitative by comparing the ground-state energy obtained
within different approximation schemes. Mixing the slave-boson scheme with
the Hartree-Fock approximation for the Hund's rule exchange has this
advantage that we can single out what feature is specifically due to the
Hund's rule coupling, as discussed next.

\subsection{Case $n=2,T=0$, and $h=0$. Effect of exchange interaction on the
metal-insulator transition (MIT).}

We consider first the simplest situation: a paramagnetic state in the
half-filled band case and at zero temperature and in zero applied magnetic
field \cite{1}.

At temperature $T=0$ the distribution function (\ref{nk}) takes the form

\begin{equation}
n_{{\bf k}}=\Theta \left( \overline{\mu }-E_{{\bf k}}\right) =\left\{ 
\begin{tabular}{ll}
$0$ & gdy $E_{{\bf k}}>\overline{\mu }$ , \\ 
$1$ & gdy $E_{{\bf k}}\leqslant \overline{\mu }$ ,
\end{tabular}
\right.  \label{nk00}
\end{equation}
where $E_k=\Phi _0\epsilon _{{\bf k}}$ and $\Phi _0\equiv q_\sigma =q_{%
\overline{\sigma }}=8d^2\left( 1-2d^2\right) .$ We obtain following
expression for the ground state energy:

\begin{eqnarray}
E_0 &=&-2\sum_\sigma \int_{-W/2}^{W/2}d\epsilon \rho \left( \epsilon \right)
\left( \overline{\mu }-\Phi _0\epsilon \right) \Theta \left( \overline{\mu }%
-\Phi _0\epsilon \right) +2Ud^2+2\bar{\mu}  \nonumber \\
&=&2\Phi _0\bar{\epsilon}+2Ud^2+K,  \label{Eo}
\end{eqnarray}
where:

\begin{equation}
\bar{\epsilon}\equiv 2\int_{-W/2}^{W/2}d\epsilon \rho \left( \epsilon
\right) \epsilon n\left( \epsilon \right)  \label{esr}
\end{equation}
is the mean bare-band energy, $n\left( \epsilon \right) =\Theta \left( 
\overline{\mu }-\Phi _0\epsilon \right) ,$ and $\rho \left( \epsilon \right) 
$ is density of states. Minimizing the functional (\ref{Eo}) with respect $%
d^2$ we obtain the solution:

\begin{equation}
d^2=\frac 14\left( 1-\frac U{8|\bar{\epsilon}|}\right) \text{,}  \label{d^2}
\end{equation}
which coincides with the results in nondegenerate-band case \cite{8,9}, i.e.
the metallic state would become unstable for $U=U_0\equiv 8\left| \bar{%
\epsilon}\right| $ if the system evolves continuously with growing $U.$
However, we shall see, that in the degenerate system this instability point $%
U_0$ is replaced by a critical point $U_c<U_0,$ at which a discontinuous
transition takes place.

Ground state energy of the metallic state can be expressed explicitly as:

\begin{equation}
E_0^M=-2|\bar{\epsilon}|-\frac 1{32}\frac U{|\bar{\epsilon}|}^2+\frac 32U-%
\frac 52J.  \label{E0}
\end{equation}

In the Mott-insulating spin-disordered state, i.e. for $d^2=0$, we have

\begin{equation}
E_0^I=K-\frac 12J=U-3J.  \label{EoI}
\end{equation}

Equating the energies (\ref{EoI}) and (\ref{Eo}) we determine the critical
value of $J=J_c$ and $U=U_c,$ for which the transition from a correlated
metal to a paramagnetic Mott insulator takes place

\begin{equation}
J_c=4|\bar{\epsilon}|\left( 1-\frac{U_c}{8|\bar{\epsilon}|}\right) ^2
\label{JU}
\end{equation}

For $J<J_c$ the metallic state is stable. Also, for $J>0$ the transition is
of the first order, whereas for $J=0$ it is continuous and corresponds to
that for nondegenerate case. For example. for $J/U=0.1$ we obtain the
critical value $U=U_c=5.1|\bar{\epsilon}|$ . The diminution of $U_c$ below $%
U_0$ is obviously caused by the Hund's rule favoring the high-spin state on
each atom.

\subsection{Case $n=2,T=0,$ and $h\geqslant 0:$ Metamagnetic transition.}

We discuss next the homogeneous magnetized state starting from the
paramagnetic state at $T=0$. From eq. (\ref{qs}) we obtain band narrowing
factor in this case in the form:

\begin{equation}
\Phi \equiv q_\sigma =\frac{16d^2}{4-m^2}\left[ 1-2d^2+\frac 12\sqrt{4\left(
1-2d^2\right) ^2-m^2}\right] .  \label{fifi}
\end{equation}

Thus, the effective mass will not depend on spin for the half-filled band
case.

The average occupancy of each orbital state with spin $\sigma $ is

\begin{equation}
n_\sigma =\frac 2N\sum_{{\bf k}}n_{{\bf k},\sigma
}=2\int_{-W/2}^{W/2}d\epsilon \rho \left( \epsilon \right) n_\sigma \left(
\epsilon \right) =2\int_{-W/2}^{(\overline{\mu }+\sigma \left( h-\beta
+Jm/4\right) )/q_\sigma }d\epsilon \rho \left( \epsilon \right) \text{ ,}
\label{ns0}
\end{equation}
where $n_\sigma \left( \epsilon \right) =\Theta \left( \overline{\mu }%
+\sigma \left( h-\beta +Jm/4\right) -q_\sigma \epsilon \right) .$ We choose
the featureless (rectangular) density of states:

\begin{equation}
\rho \left( \epsilon \right) =\left\{ 
\begin{tabular}{ll}
$\frac 1W$ & for $\epsilon \in \left[ -W/2,W/2\right] $ , \\ 
$0$ & otherwise.
\end{tabular}
\right.  \label{1/W}
\end{equation}

The overall futures should be independent of the detailed shape of $\rho
\left( \epsilon \right) ,$ as we determine global quantities ($m,$ $E_G,$ $%
d^2$) involving integrals over the filled part of the relevant bands.
Putting above DOS into (\ref{ns0}) we obtain:

\begin{equation}
n_\sigma =\frac 2W\int_{-W/2}^{(\overline{\mu }+\sigma \left( h-\beta
+Jm/4\right) )/\Phi }d\epsilon =\frac 1{W\Phi }\left[ \left( h-\beta +\frac 1%
4Jm\right) \sigma +\bar{\mu}+\frac{W\Phi }2\right] \text{ .}  \label{ns1/w}
\end{equation}

Summation over $\sigma $ of the above equation leads to the condition $\bar{%
\mu}=0$, but multiplying it by $\sigma $ and then summing over $\sigma ,$
leads to the expression for the molecular field of the form

\begin{equation}
\beta =-\frac 14\Phi Wm+h+\frac 14Jm\text{ .}  \label{beta}
\end{equation}
Because of $\Phi $ is a nonlinear in magnetic moment $m,$ the molecular
field $\beta $ depends also on $m$ in the same manner. Inserting $\beta $
into (\ref{F/2N}) for $T=0,$ we obtain: 
\begin{equation}
E_M\equiv \frac F{NW}=-\left( 4-m^2\right) \Phi /8+2ud^2-hm-\frac 18jm^2+k,
\label{Eg}
\end{equation}
where the reduced parameters are:

\begin{equation}
u=U/W,\text{ }j=J/W,\text{ }k=K/W,\text{ }h\rightarrow h/W.  \label{all/W}
\end{equation}

We determine the saddle-point solution minimizing $E_M$ with respect $d^2$
and $m.$ Differentiation with respect to $m$ leads to

\begin{equation}
d^4\left( m^2-16a\right) +a\left( 16d^2+m^2-4\right) =0\text{ .}  \label{dEm}
\end{equation}
where $a=\left( h+jm/4\right) ^2.$ The differentiation with respect $d^2$
yields

\begin{equation}
\frac 12\left( u+4d^2-1\right) \sqrt{4\left( 1-2d^2\right) ^2-m^2}-\left(
1-2d^2\right) ^2+\frac 14m^2+2d^2\left( 1-2d^2\right) =0\text{ .}
\label{dEd}
\end{equation}

For $h=0$ and $u\leqslant 2$ from (\ref{d^2}) we obtain $d^2=(1-u/2)/4$. The
change of direction of magnetic field $h$ does not change $d^2$; thus $%
(\partial d^2/\partial h)_{h=0}=0.$ Next, differentiating with respect $h$
eq. (\ref{dEm}) and taking into account the above relations we obtain the
following expression for the magnetic susceptibility:

\begin{equation}
\chi \left( 0\right) \equiv \left. \frac{\partial m}{\partial h}\right|
_{h=0}\frac 14\chi _0=\frac{\chi _0}{\frac{2-u}{2+u}-j}\text{ ,}
\label{hihihi}
\end{equation}
where $\chi _0=2\mu _B^2/W$ is the Pauli susceptibility (cf. Appendix A).
The magnetic susceptibility was derived for the metallic state, so from
taking into account the condition (\ref{JU}) we obtain the metallic state
for $j<(1-u/2)^2\leqslant \frac{1-u/2}{1+u/2}\leqslant 1.$ From, the last
inequality we obtain the condition for the magnetization in small fields:

\begin{equation}
m\approx \left. \frac{\partial m}{\partial h}\right| _{h=0}h=\frac{4h}{\frac{%
2-u}{2+u}-j}>\frac{4h}{1-j}\text{ .}  \label{m2hj}
\end{equation}

This condition reduces solutions of eq. (\ref{dEm}) to:

\begin{equation}
d^2=\frac{-8a+m\sqrt{a\left( 16a+4-m^2\right) }}{m^2-16a}\text{ ,}
\label{d2m}
\end{equation}

Substituting $d^2$ into eq. (\ref{dEd}) we obtain following equation for
magnetization:

\begin{eqnarray}
&&-64a^3+12a^2m^2-3am^4/4+m^6/64+64a^2u-m^4u/4+  \nonumber \\
&&-16au^2+32a^2u^2+m^2u^2+12am^2u^2+m^4u^2/8+  \nonumber \\
&&-16au^3-m^2u^3-4au^4+m^2u^4/4=0.  \label{rownm}
\end{eqnarray}

The pair of solutions $\left( m,d^2\right) $ determines the ground state
energy (\ref{Eg}). Ground state energy of a Mott-Hubbard insulator in the
spin disordered state is:

\begin{equation}
E_I=-2h-j/2+k.  \label{EIh}
\end{equation}

From the preceding subsection we know that the metal-insulator transition
take place for $j=\left( 1-u/2\right) ^2$. From the expression (\ref{hihihi}%
) for the magnetic susceptibility we see that except for the case $j=0$ and $%
u=2,${\it \ there is no singularity in }$\chi $ {\it at the transition}.
This means that the transition for $j>0$ is of the first-order$.$ For $j=0$
we have second order transition, as in the nondegenerate band case, where $%
\chi $ is singular (cf. Appendix A).

Field dependences of $d^2$ and $m$ is displayed on Figs.1ab, respectively.
Close to transition point the system exhibits a metamagnetic behavior i.e.
the $m\left( h\right) $ dependence in curved upwards. With the increasing
field $h$ the doubly occupancy diminishes because of the growing spin
polarization and at the transition point $d^2$ jumps to zero, while the
magnetic moment jumps to its saturation value. The magnitude of
magnetization jump $\Delta m$ and the critical field $h_c$ depend on the
values of $j$ and $u$, as shown in Fig.2. One observes here an example of a
magnetic-field-induced localization, with a formation of a ferromagnetic
(field-induced) insulator. The transition from the metal to the insulator
can be either first or second kind, depending on values $u$ and $j.$ Regimes
of $u$ and $j,$ where the particular kind of a transition takes place, are
displayed in Fig.3. We mark the area of insulating phase in zero field
bounded by the curve determined from Eq. (\ref{JU}). Insulator polarizes at $%
T=0$ to a saturated state in an infinitesimal field; jump of the
magnetization is then equal to $\Delta m=2.$ Upon increasing of $j$ the
critical field $h_c$ decreases; likewise the magnetization jump. From Fig.2
and Fig.3 one could seen that for $u>0.5$ a second-order transition does not
occur. Obviously, our analysis is valid quantitatively only for $j$
substantially smaller than $u.$ Also, the critical field for metamagnetic
transition becomes experimentally accessible only close to MIT ($%
u\rightarrow 2$) and reduces gradually to zero with growing $j.$ This
prediction could be tested experimentally (see below).

\subsection{Case $n<2$ and $T=0$: spin-split masses.}

Consider now briefly the situation $n<2$ when the metallic state is stable.
We examine the effect of magnetic field on the transition to the saturated
state, as a function of the parameters $u$ and $j.$ Ground state energy for
a polarized state is provided by the expression:

\begin{equation}
E_M\left( n\right) \equiv \frac F{NW}=-\frac 1{16}\sum_\sigma q_\sigma
\left( 4-n-\sigma m\right) \left( n+\sigma m\right) +2ud^2-\frac 18jm^2-hm+%
\frac 14kn^2.  \label{F/N_}
\end{equation}

Energy of the saturated state is obtained by taking into account eq. (\ref
{qs}) and making substitution $d^2\rightarrow 0$ and $m\rightarrow n$ in (%
\ref{F/N_}). We have then:

\begin{equation}
E_I\left( n\right) \equiv -\frac 12n\left( 2-n\right) -\frac 18jn^2-hn+\frac %
14kn^2.  \label{Eo(n)}
\end{equation}

In Fig.4 we display magnetization $m$, doubly occupancy $d^2$ and the mass
enhancement $q_\sigma ^{-1}$ as a function of applied magnetic field for $%
u=1.98$ and $j=0.05$, for the band fillings $n=1.86$ and $n=1.8$ . The
important feature is that the mass enhancement is spin dependent (see the
lowest panel). Also, there is a possibility of first and second order
transitions to a saturated state. Regimes of $u$ and $j$ , where this two
types of transition are possible for filling $n=1.8$ are shown in Fig.5. By
comparing this Figure with Fig.3 we see that regime of stability of an
insulator for $n=2$ corresponds here to the area Ib with excluded doubly
occupancy (EDO), when the system is in a magnetically saturated state in a
vanishing field. In the case of one-band model, i.e. $j=0,$ EDO state is
reached for $u\rightarrow +\infty $ in $h=0.$ For $j\neq 0,$ the doubly
degenerate case, this state is achieved for finite $u.$

One should note that the effect of spin-split masses arising in the
magnetically polarized state is not associated with the emergence of
spin-dependent density of states in the bare band. This can be seen from the
fact that the quasiparticle energies $E_{{\bf k}\sigma }$ (cf. Eq.(24)) lead
to the following density of states: 
\begin{equation}
\rho _\sigma \left( E\right) =\frac 1{q_\sigma }\rho _\sigma ^{\left(
0\right) }\left( \varepsilon -\frac{\sigma \left( h+\frac 14Jm-\beta \right) 
}{q_\sigma }\right) ,
\end{equation}
where $\rho _\sigma ^{\left( 0\right) }\left( \varepsilon \right) $ is the
density of bare states per spin. Thus, the enhancement due to the
correlations is distinct from the change of the density of states caused by
the presence of the effective field. The many-body nature of the enhancement
factor can be also seen by writing the quasiparticle energy in the form 
\begin{equation}
E_{{\bf k}\sigma }=\varepsilon _{{\bf k}}-\sigma h+\left( q_\sigma -1\right)
\varepsilon _{{\bf k}}-\sigma \left( \frac 14Jm-\beta \right) \equiv
\varepsilon _{{\bf k}}-\sigma h-\Sigma _\sigma \left( \omega =\varepsilon _{%
{\bf k}}\right) ,
\end{equation}
where the selfenergy is 
\begin{equation}
\Sigma _\sigma \left( \omega \right) =-\left( 1-q_\sigma \right) \omega
-\sigma \left( \frac 14Jm-\beta \right) .
\end{equation}

Taking into account the well-known definition of the mass enhancement in the
Fermi liquid 
\begin{equation}
\frac{m_0}{m^{*}}=\lim_{\omega \rightarrow \mu }\left( 1+{\frac \partial {%
\partial \omega }}\mathop{\rm Re}\Sigma _\sigma \left( \omega \right)
\right) ,
\end{equation}
we have that $m_0/m^{*}=q_\sigma ,$ i.e. is indeed spin dependent. This spin
dependence, which will lead to the strong field dependence of the linear
specific heat close to the metamagnetic transition (see Spa\l ek at al. \cite
{11}) is characteristic of the almost localized fermions and should be
determined experimentally.

\section{Mott-Hubbard boundary at nonzero temperature}

\subsection{The phase diagram: $h\geqslant 0.$}

In the preceding Section we considered the system properties at zero
temperature. We concentrate now on a more realistic case of non-zero
temperature in the half-filled-band case. The low temperature (Sommerfeld)
expansion \cite{9} of the free energy (\ref{F/2N}) for the metallic phase
leads to the following expression for the constant DOS: 
\begin{equation}
\frac{F_M}{NW}=-\left( 4-m^2\right) \Phi /8+2ud^2-hm-\frac 18jm^2+k-\frac{%
2\pi ^2}{3\Phi }\tau ^2,  \label{F(T)}
\end{equation}
where $\tau =k_BT/W.$ No higher-order term in $\tau $ appears for this DOS.

Hamiltonian describing the insulating phase can be rewritten in the form

\begin{eqnarray}
{\cal H}_I &=&K\sum_in_{i1}n_{i2}-2J\sum_i{\bf S}_{i1}\cdot {\bf S}%
_{i2}-2h\sum_{il}S_{il}^z\ =  \nonumber \\
&=&\sum_i\left[ Kn_{i1}n_{i2}-J\left( {\bf S}_i^2-{\bf S}_{i1}^2-{\bf S}%
_{i2}^2\right) -2hS_i^z\right] .  \label{HI}
\end{eqnarray}
where ${\bf S}_i={\bf S}_{i1}+{\bf S}_{i2}$ is the total spin per site. The
eigenstates of $H_I$ are singlet and triplet configurations. The partition
function is then given by:

\begin{eqnarray}
Z &=&Tr\left( e^{-\beta {\cal H}_I}\right) =e^{-\beta KN}\left[ e^{-3\beta
J/2}+e^{\beta J/2}+e^{\beta J/2}\left( e^{2\beta h}+e^{-2\beta h}\right)
\right] ^N=  \nonumber \\
&=&e^{-\beta KN}\left[ e^{-3\beta J/2}+e^{\beta J/2}\left( 1+2\cosh \left(
2\beta h\right) \right) \right] ^N.  \label{ZI}
\end{eqnarray}

This expression leads to the free energy for the Mott insulating state in
the following form

\begin{eqnarray}
\frac{F_I}{NW} &=&-\frac 1{NW\beta }\ln \left[ Z\right] =k-\frac 1{W\beta }%
\ln \left[ e^{-3\beta J/2}+e^{\beta J/2}\left( 1+2\cosh \left( 2\beta
h\right) \right) \right] =  \nonumber \\
&=&k-\tau \ln \left[ e^{-3j/2\tau }+e^{j/2\tau }\left( 1+2\cosh \left(
2h/\tau \right) \right) \right] .  \label{FI}
\end{eqnarray}

The result in the limit $\tau \rightarrow 0$ reduces to the expression (\ref
{EIh}). The magnetization in this state is

\begin{eqnarray}
m &\equiv &2Tr\left( S_i^ze^{-\beta {\cal H}_I}\right) =2\frac{e^{-\beta
KN}e^{\beta J/2}\left( e^{2\beta h}-e^{-2\beta h}\right) }{e^{-\beta
KN}\left[ e^{-3\beta J/2}+e^{\beta J/2}\left( 1+2\cosh \left( 2\beta
h\right) \right) \right] }=  \nonumber \\
&=&\frac{4\sinh \left( 2\beta h\right) }{e^{-2J\beta }+1+2\cosh \left(
2\beta h\right) }=\tanh \left( h/\tau \right) \frac 2{\frac{e^{-2j/\tau }-1}{%
4\cosh ^2\left( h/\tau \right) }+1},  \label{mI}
\end{eqnarray}

For $j=0$ we obtain $m=2\tanh \left( h/\tau \right) $, the expression for
the magnetization of non-interacting spins. Such a situation arises because
the saddle-point approximation in its essence is a single-site approximation
(the intersite interaction arises from the quantum Gaussian fluctuations
around it).

The field dependences of the magnetization and of the doubly occupancy are
shown in Fig.6. We see that upon increasing $j$ while keeping fixed $u$ the
critical metamagnetic field is reduced. It indicates, as before for $T=0$, a
significant role of exchange interaction and associated with it Hund's rule.
Also, the magnetization curve is slightly curved upwards in small fields.
The upper part of the $m(h)$ curve reflects magnetization of the
localized-moment system. Thus we have a transition from an itinerant (albeit
metamagnetic) to localized-type behavior as a function of $h.$

Substituting the expressions for $m$ and $d^2$ taken from Appendix A into
the free energy (\ref{F(T)}) we can determine the system behavior in the
vicinity of $h=0.$ The phase diagram for cases $h=0$ and $0.01$ , for
different values of $J/U$ is exhibited in Fig.7. Upon increasing of $J/U$
the paramagnetic insulating (PI) phase expands at the expense of the
paramagnetic metallic (PM) phase . In the applied field the boundaries shift
towards lower temperatures. We see also a typical reentrant metallic
behavior at high temperatures. Namely, with rising temperature the system
evolves from a metal through an insulator back to the metallic state. In the
inset we display the temperature dependence of the free energies for PM and
PI states.

The shape of the phase boundary is essentially the same as in
nondegenerate-band case \cite{9}. However, it is shifted remarkably towards
lower values of $U/W$ already for rather small values of $J/W.$ The boundary
is of first order apart from the point specified. The upper part of the
curve is only qualitative, particularly if realistic DOS is used, as
higher-order contribution in $\tau $ will become important. In fact, the
solution of the Hubbard model in infinite-dimension limit \cite{10} provides
only a crossover behavior, not a weakly discontinuous retrograde behavior.

One may ask if the low-$T$ analysis is realistic, since in the mean-field
slave-boson analysis the low-energy spin fluctuations (quantum Gaussian
fluctuations around the saddle-point here) have been neglected. Those spin
fluctuations lead to the contribution $\sim T^4\ln \left( T/\theta \right) $
in the free energy \cite{14}. This contribution is of higher order than the $%
T^2$ contribution coming from the quasiparticle excitations across the Fermi
surface. Therefore, the analysis is realistic, but only to the leading order
in low-$T$ expansion, as we have done it.

\subsection{Physical discussion}

The evolution of the almost localized fermions discussed above for the
half-filled band case can be explained nicely from a physical point of view.
Namely, for $U\rightarrow U_c$ the renormalized band ($\Phi \bar{\epsilon}$)
and the correlation ($Ud^2$) energies {\it almost compensate each other. }In
such situation, much smaller entropy ($\sim TS$) or applied field ($\sim \mu
_BH_a$) energies tip the balance towards either M or I phase. Explicitly, at
low temperatures, the spin-disordered magnetic insulator has much larger
entropy contribution ($-k_BT\ln 2,$ per orbital) than that of the almost
localized Fermi liquid ($-\gamma _0T^2/\left( 2\Phi \right) $ ,with $\gamma
_0=\left( 2/3\right) \pi ^2k_B^2\rho $). This circumstance tips the balance
from PM state (with $E_G<0,$ but small entropy contribution) towards PI
state (with $E_G=0,$ but much larger entropy contribution). At much higher
temperatures the balance is tipped back towards PM phase, since eventually
the entropy of the metallic state grows and approaches the asymptotic value $%
2k_B\ln 2$ per orbital. Thus the reentrant metallic behavior is driven by
the entropy. It is observed in both (V$_{1-x}$Cr$_x$)$_2$O$_3$ \cite{15} and
NiS$_{2-x}$Se$_x$ \cite{3} systems. It can be applied also to explain the
low-temperature reentrant liquid behavior in liquid $^3$He \cite{16}. In our
view the reentrant behavior appearing either as a crossover or as a
discontinuous transition ($^3$He) uniquely present in the Mott-Hubbard
systems defined as systems, for which the band and the Coulomb parts of
their energy almost compensate each other. The exchange contribution tips
the balance further towards the localized state.

Obviously, the first-order nature of the transition will lead to the
coexistence of the two phases, with localized and itinerant electrons,
respectively. However, this mixed phase can be discussed only when the
magnetism is included, and we will not elaborate on it here.

As mentioned at the beginning, the present formulation represents for $%
H_a=T=0$ a simplified version of the full slave-boson and Gutzwiller
treatments \cite{1} discussed recently. In Fig.8 we have compared the
Hartree-Fock (HF), ours (SB-HF) and full slave-boson (SB) results for the
ground state energies for $h=0$ and $n=2.$ the arrows denote the position of
the Mott-Hubbard boundary in the two latter schemes. The energy difference
diminishes rapidly with growing $h.$ The difference for physical quantities
is only quantitative, not qualitative, though the SB approach \cite{1} has
slightly lower energy, since it contains many more variational parameters.
In the next Section we extend the main features of our solution to arbitrary 
$J/W$ and higher degeneracy $d$ of equivalent orbitals and an arbitrary
filling.

\section{Metal-insulator transition for arbitrary orbital degeneracy and
filling: a spin rotation invariant model}

\subsection{Global (site) representation of the intraatomic interaction.}

We now generalize the principal features of our argument to the case of
orbital degeneracy. First, we represent the intraatomic part in terms of
global (site) representation. For that purpose we start from the following
expression of that part

\begin{equation}
{\cal H}_I=U\sum_{il}n_{il\uparrow }n_{il\downarrow }+\frac 12%
K\sum\limits_{ill^{^{\prime }}\sigma \sigma ^{^{\prime }}}{}^{^{\prime
}}n_{il\sigma }n_{il^{^{\prime }}\sigma ^{^{\prime }}}-J\sum_{ill^{^{\prime
}}}{}^{^{\prime }}{\bf S}_{il}\cdot {\bf S}_{il^{^{\prime }}}\text{ ,}
\label{1}
\end{equation}
where now $l$ and $l^{^{\prime }}$ assume the values $1,2,\ldots ,d$ , and
the primed summation is taken for $l\neq l^{^{\prime }}.$ We introduce the
global spin and particle number operators through the relations

\begin{equation}
{\bf S}_i\equiv \left( S_i^{+},S_i^{-},S_i^z\right) \equiv \sum_{l=1}^d{\bf S%
}_{il}\text{ ,}  \label{2}
\end{equation}
and

\begin{equation}
n_i\equiv \sum_\sigma n_{i\sigma }\equiv \sum_{l\sigma }n_{il\sigma }\text{ .%
}  \label{3}
\end{equation}

We have the following relations between global and previously introduced
operators

\begin{equation}
\sum_{ll^{^{\prime }}}n_{il}n_{il^{^{\prime
}}}=n_i^2-n_i-2\sum_ln_{il\uparrow }n_{il\downarrow }\text{ ,}  \label{4}
\end{equation}
and 
\begin{equation}
\sum_{ll^{^{\prime }}}{}^{^{\prime }}{\bf S}_{il}\cdot {\bf S}_{il^{^{\prime
}}}=S_i^z+\frac 32\sum_ln_{il\uparrow }n_{il\downarrow }-\frac 34n_i\text{ .}
\label{5}
\end{equation}

Hence, up to constant ${\cal H}_I$ takes the form

\begin{equation}
{\cal H}_I=\frac 12K\sum_in_i^2-J\sum_i{\bf S}_i^2+I\sum_{il}n_{il\uparrow
}n_{il\downarrow },  \label{6}
\end{equation}
where $I=U-K-\frac 32J.$

For particular example of $e_g$ band we have

\begin{equation}
{\cal H}_I=\frac 12\left( U-\frac 52J\right) \sum_in_i^2-J\sum_i{\bf S}%
_i^2+J\sum_{il}n_{il\uparrow }n_{il\downarrow }.  \label{7}
\end{equation}

The first term represents charge fluctuations, the second the atomic Hund's
rule. Moreover, unlike in a nondegenerate band the intraorbital intraatomic
(Hubbard) parameter ($J$) is much smaller than $U.$ Additionally, the first
two terms are proportional to $d^2$ while the third is $\sim d.$ Hence, the
last term can become the smallest, particularly for highly degenerate
systems. In general, the Hamiltonian (\ref{6}) expresses the so-called
minimum polarity model of Van Vleck (charge fluctuations suppressed by
growing $U$), as well as the separation of the dynamic processes into
inequivalent charge and spin degrees of freedom in the effective one-band
model introduced by Hubbard \cite{17}. For highly degenerate systems we can
approximate

\begin{equation}
I\sum_{il}n_{il\uparrow }n_{il\downarrow }\simeq I\sum_{il}\frac{%
n_{i\uparrow }}d\cdot \frac{n_{i\downarrow }}d=\frac Id\sum_in_{i\uparrow
}n_{i\downarrow }\text{ }.  \label{8}
\end{equation}

Noting that

\begin{equation}
n_{i\uparrow }n_{i\downarrow }=\frac{n_i^2}4-\left( {\bf \mu }_i\cdot {\bf S}%
_i\right) ^2,  \label{9}
\end{equation}
where ${\bf \mu }_i$ is unit vector along an arbitrarily oriented spin
quantization axis for the spin ${\bf S}_i,$ one has finally

\begin{equation}
{\cal H}_I\simeq \frac{\tilde{U}}4\sum_in_i^2-J\sum_i{\bf S}_i^2-\frac Id%
\sum_i\left( {\bf \mu }_i\cdot {\bf S}_i\right) ^2,
\end{equation}
where $\tilde{U}\equiv 2K+I/d.$ For a particular case of a nondegenerate
band $(d=1)$, $K=J=0$ we recover the earlier results \cite{18} with $\tilde{U%
}=U,$ and $I=U.$ For $d\rightarrow \infty $ the last term is absent and the
correlated paramagnetic state at $T=0$ is described by two variational
parameters:

\begin{equation}
\lambda \equiv \langle n_i^2\rangle ,
\end{equation}
and 
\begin{equation}
m\equiv \langle {\bf S}_i^2\rangle .
\end{equation}

Note that in this Section $m$ represent the local-moment magnitude. The
third parameter describing the long-range order is obtained by making the
Hartree-Fock approximation 
\begin{equation}
\frac Id\sum_i\left( {\bf \mu }_i\cdot {\bf S}_i\right) ^2\simeq \frac Id%
\sum_i\left( {\bf \mu }_i\cdot \langle {\bf S}_i\rangle \right) {\bf \mu }%
_i\cdot {\bf S}_i\text{ .}
\end{equation}

The collinear magnetic ordering is expressed then through $\langle
S_i^z\rangle ={\bf \mu }_i\cdot \langle {\bf S}_i\rangle .$ Since in this
Section we consider only a paramagnetic state at $T=H_a=0,$ we neglect in
our analysis the last term. Note however, that the present parameters $%
\lambda ,$ $m$ and $\langle {\bf \mu }_i\cdot {\bf S}_i\rangle $ for
arbitrary $d$ correspond directly to the parameters $d^2,$ $p_{\uparrow }^2$
and $p_{\downarrow }^2$ in the preceding Section ($e^2$ is removed via the
completeness condition $e^2+p_{\uparrow }^2+p_{\downarrow }^2+d^2=1).$

\subsection{Magnitude of local-moment and charge fluctuations: first-order
transition to the insulating state at $T=0{\bf .}$}

In direct analogy to the doubly degenerate case discussed in Section II, we
express the ground state energy in the form

\begin{equation}
\frac{E_G}N=\Phi \left( \lambda ,m\right) \bar{\epsilon}+\frac{\tilde{U}}4%
\lambda -Jm\text{ ,}  \label{SpEg}
\end{equation}
where now the band narrowing factor $\Phi $ depends on $\lambda $ and $\bar{%
\epsilon}$ represent the average band energy for a degenerate system.
Without a loss of generality one can assume that $\Phi (\lambda ,m)=\Lambda
\left( m\right) G\left( m\right) .$ In accordance with our simple derivation
of the mean-field (Gutzwiller) approach in the half-filled band case \cite
{19} we make an expansion:

\begin{equation}
\Lambda \left( \lambda \right) =l_0+l_1\lambda +l_2\lambda ^2,
\end{equation}
and 
\begin{equation}
G\left( m\right) =g_0+g_1m+g_2m^2.
\end{equation}

Note that $0\leqslant \Phi \equiv \Lambda G\leqslant 1.$ The expansion has
the meaning of a Landau expansion, and the coefficients can be determined by
calculating $E_G$ explicitly in limiting situations. For example, in the
Hartree-Fock approximation we have $\Lambda =G=1$ and thus elementary
analysis provides us with

\begin{equation}
\lambda \equiv \lambda _0=n+n^2\left( 1-\frac 1{2d}\right) ,
\end{equation}
and 
\begin{equation}
m\equiv m_0=\frac 34n\left( 1-\frac n{2d}\right) ,
\end{equation}
where $n$ is the band filling ($n=d$ corresponds to the half-filling).
Analogously, for $\tilde{U}\rightarrow \infty $ and $J\rightarrow \infty ,$
the quantities $\lambda _0$ and $m_0$ will approach their atomic values $%
\lambda _\infty ,$ $m_\infty $

\begin{equation}
m\equiv m_\infty =\left\{ 
\begin{tabular}{cc}
$\frac 34n$ & for $n\leqslant 1$ \\ 
$\frac n2\left( \frac n2+1\right) $ & for $n\geqslant 1$,
\end{tabular}
\right.
\end{equation}
and 
\begin{equation}
\lambda \equiv \lambda _\infty =\left\{ 
\begin{tabular}{cc}
$n$ & for $n\leqslant 1$ \\ 
$n^2$ & for $n\geqslant 1$.
\end{tabular}
\right.
\end{equation}

Additionally, applying the equilibrium conditions

\begin{equation}
\frac{\partial E_G}{\partial \lambda _0}=\frac{\partial E_G}{\partial m_0}=0,
\end{equation}
we obtain the expression

\begin{eqnarray}
\lambda &=&\lambda _0\left[ 1-\frac{\tilde{U}}{8l_2\lambda _0\left| \bar{%
\epsilon}\right| G\left( m\right) }\right] , \\
m &=&m_0\left[ 1+\frac J{2g_2m_0\left| \bar{\epsilon}\right| \Lambda \left(
\lambda \right) }\right] .
\end{eqnarray}

As in nondegenerate case, the charge fluctuations are suppressed with
increasing $\tilde{U}.$ On the contrary, the magnetic moment grows with
increasing $J.$ Additionally, using the conditions $\Lambda \left( \lambda
_0\right) =G\left( m_0\right) =1$ we obtain that

\begin{eqnarray}
\Lambda \left( \lambda \right) &=&1+l_2\left( \lambda -\lambda _0\right) ^2,
\\
G\left( m\right) &=&1+g_2\left( m-m_0\right) ^2.
\end{eqnarray}

The coefficients $l_2$ and $g_2$ are determined from the condition that for
any integer band filling $n\geqslant 1$ we have that $\Lambda \left( \lambda
_\infty \right) =0,$ and $G\left( m_\infty \right) =n_\sigma /n=1/2$ (this
result is valid for the paramagnetic configuration only and reflects the
Pauli exclusion). So, finally we have

\begin{equation}
\Lambda \left( \lambda \right) =1-\left( \frac{\lambda -\lambda _0}{\lambda
_\infty -\lambda _0}\right) ^2,
\end{equation}
and 
\begin{equation}
G\left( m\right) =1-\frac 12\left( \frac{m-m_0}{m_\infty -m_0}\right) ^2.
\end{equation}

Substituting these expressions to the expression (\ref{SpEg}) for $E_G,$ we
obtain the explicit form in the terms of variational variables $\lambda $
and $m.$ Making use of the conditions $\partial E_G/\partial m=\partial
E_G/\partial \lambda =0$ we arrive at the algebraic equations, which may be
transformed into the following equations for the functions $G$ and $\Lambda $
taken at the extremal points:

\begin{equation}
G^3-G^2\left[ 1-2\frac{\tilde{U}}{U_0}+\left( \frac J{J_c}\right) ^2\frac 1{%
g_2}\right] +G\frac{\tilde{U}}{U_0}\left( \frac{\tilde{U}}{U_0}-2\right)
-\left( \frac{\tilde{U}}{U_0}\right) ^2=0,
\end{equation}
and 
\begin{equation}
\Lambda ^3-\Lambda ^2\left[ 1+2\frac J{J_c}+\left( \frac{\tilde{U}}{U_0}%
\right) ^2\frac 1{l_2}\right] +\Lambda \frac J{J_c}\left( \frac J{J_c}%
+2\right) -\left( \frac J{J_c}\right) ^2=0,
\end{equation}
with $U_0=8\left| \bar{\epsilon}\right| ,$ and $J_c=2\left| \bar{\epsilon}%
\right| .$ These two equations can be transformed into each other by changes 
$\Lambda \leftrightarrow G,$ $g_2\leftrightarrow l_2,$ and $%
J/J_c\leftrightarrow -\tilde{U}/U_0.$ Hence, it is sufficient to solve
numerically one of them and adapt it subsequently for the second equation.
One should notice that the nondegenerate-band-case result $\lambda =\lambda
_\infty $ (i.e. $d^2=0$) is recovered for $U/U_0=1.$

Before going into the numerical analysis let us summarize the above
Subsection. We have developed a relatively simple scheme of calculating the
magnitudes of local moment $\langle {\bf S}_i^2\rangle $ and of the charge
fluctuations $\langle n_i^2\rangle ,$ which is equivalent to
Gutzwiller-Brinkman-Rice scheme \cite{8} for $d=1.$ These magnitudes are
calculated from a relative balance between the renormalized band energy from
one side, and the correlation energies ($\frac{\tilde{U}}4\lambda ^2-Jm^2$)
from the other. The method involves an interpolation between low- and
high-correlation regimes (note that the mean-field slave boson theory
requires such an interpolation to the $U\rightarrow 0$ limit \cite{13,12}).

\subsection{Numerical analysis for arbitrary degeneracy and filling}

We define the reduced variables

\begin{equation}
\lambda _R=\frac{\lambda -\lambda _0}{\lambda _\infty -\lambda _0},
\end{equation}
and 
\begin{equation}
m_R=\frac{m-m_0}{m_\infty -m_0}.
\end{equation}

In this manner, the values $\lambda _R=m_R=0$ corresponds to the
Hartree-Fock approximation, whereas the limit with $\lambda _R=m_R=1$
correspond to the exact atomic limit, which for an integer $n\geqslant 1$
corresponds to the Mott-Hubbard insulator. In Fig.9 we have plotted $E_G$ as
a function of $U_R\equiv \tilde{U}/U_0=\tilde{U}/8\left| \bar{\epsilon}%
\right| ,$ for $J_R/U_R=0.1,$ which corresponds to $J/\tilde{U}=0.4.$ The
bare band energy for the featureless density of states is $\bar{\epsilon}%
=-\left( W/2\right) n\left( 1-n/2d\right) .$ One sees that $E_G=0$ for a
critical value of $U_R.$ At this point the values $\lambda _\infty $ and $%
m_\infty $ are reached for $n>1$ in a discontinuous way, as illustrated in
Figs.10ab. The localization threshold diminishes with $n.$ This is because
the band energy varies roughly $\sim n,$ while the interaction energy is $%
\sim n^2.$ The transition to the localized-moment state is determined by the
interplay between the exchange and the Coulomb interaction. This is
illustrated in Figs.11ab, where the band-narrowing factors $\Lambda $ and $G$
have been specified. In most situations the part $G\left( m\right) $ changes
from its Hartree-Fock value only a little. This provides an a posteriori
justification of the Hartree-Fock approximation for the exchange term. The
dominant role of the term $\sim n_i^2$ over the Hund's rule term gives some
support to the interpretation of the local-moment formation in the terms of
a nondegenerate Hubbard model. This becomes clear if one notices that in
that case

\begin{equation}
\lambda =\langle n_i^2\rangle =n+2\langle n_{i\uparrow }n_{i\downarrow
}\rangle =n+2d^2,
\end{equation}
and hence $\lambda =\lambda _\infty =n$ correspond to the limit $d^2=0$ ($%
\Lambda =0$). In the situation depicted in Fig.11a $\Lambda \approx 0.6$ at
the transition. However, unlike in nondegenerate system, the spin-charge
fluctuation coupling is leading to the discontinuous character of this
transition for $n>1.$

The discontinuous nature of the metal-insulator transition induces only a
weak enhancement of the effective mass close to the transition (the curves
in Fig.10b are physically meaningful only below the discontinuity points).
The results in this respect are valid universally for arbitrary $d>1$ and $%
n>1.$

\section{Conclusions}

In this paper we have put an emphasis on the similarities and differences of
the Mott localization in degenerate-band system with the extensive analysis
of MIT existing for the nondegenerate band case. For that purpose we have
made in the first part the Hartree-Fock approximation for the exchange
interaction. This approximation is applicable in the limit when $J$ is
substantially smaller than $U,$ as is usually the case for $3d$ bands. This
scheme provides us with the physically plausible conclusion that the
metal-insulator transition is mainly driven by the intraatomic Coulomb
interaction; the intraatomic exchange is responsible for the first-order
nature of the transition already at $T=0.$ Also, with this approach one can
see that the correlated systems, whether orbitally degenerate or not, can in
the mean-field approximation be described as systems of fermionic
quasiparticles with spin-split masses (in the magnetically polarized state)
and a nonlinear molecular field coming from the correlations, in addition to
the usual exchange field.

The fundamental question is whether the present approach (as well as those
listed in Ref.1) provide a proper mean-field theory of a correlated state
near the metal-insulator transition. It seems so and the proper order
parameter in the orbitally nondegenerate paramagnetic system is either $\Phi 
$ or $d^2$, which are nonzero in the metallic phase and vanish in the
insulating state. What is more important, since for $n=2$ the band-narrowing
factor $\Phi $ in PM state can be directly related \cite{20} to the physical
quantity $Z(=\Phi )$ representing the discontinuity of the Fermi-Dirac
distribution at the Fermi energy, the order parameter is a measurable
quantity for this quantum phase transition at $T=0.$ In the magnetically
ordered state the system is additionally characterized by a staggered moment 
$\langle S_i^z\rangle .$ In the second part of this paper we have shown that
even in the paramagnetic state the orbitally degenerate system is
additionally characterized by the parameter $\langle {\bf S}_i^2\rangle $
describing the local moment magnitude. Thus, in our view, the set of the
parameters: $d^2\Leftrightarrow \langle n_i^2\rangle ,$ $\langle {\bf S}%
_i^2\rangle ,$ and ${\bf \mu }_i\cdot \langle {\bf S}_i\rangle $ compose a 
{\it minimal set }describing the metal-insulator transition in the
degenerate system and associated with it magnetic transition (in
nondegenerate system $\langle {\bf S}_i^2\rangle =\frac 34\left(
1-2d^2\right) $). Obviously, the full slave boson and Gutzwiller approaches 
\cite{1} provide an essentially the same qualitative picture, although
contain more parameters, which are eliminated by implementing the
constraints appearing as consistency conditions. In this respect, our
simplified approach provides a didactical guidance for more complicated
analysis. For example, the effect of quantum Gaussian fluctuations in
auxiliary Bose fields neglected so far will introduce intersite exchange
interactions, which will not introduce any additional order parameter,
though the detailed thermodynamic properties will contain the contribution
coming from the interaction between the quasiparticles with the novel
characteristics ($q_\sigma ,$ $\beta ,$ $\beta _0$), which appeared on the
mean-field level.

The slave boson approach has been recently extended \cite{22} to describe
the antiferromagnetic phase in the doubly degenerate Hubbard model. However,
the temperature dependence of MIT in the paramagnetic phase is still
important, since in NiS$_{2-x}$Se$_x$ \cite{3} one observes a transition to
the semiconducting phase upon heating the system, when system crosses the
N\'{e}el point (cf. Fig.7 for $H_a=0$, where M$\rightarrow $I transition is
observed upon heating the system!). The detailed analysis requires the
discussion of antiferromagnetic insulating and metallic states at $T>0$
before any direct comparison with the experiment is made.

\begin{center}
\ \noindent
{\bf {ACKNOWLEDGMENTS} }
\end{center}

The authors are grateful to Prof. J. M. Honig and Dr. Mireille Lavagna for
discussions. The work was supported in part by the KBN Grant No. 2PO3B 129
12, and in part by NSF Grant No. DMR 96 12130.

\begin{center}
\newpage \ \noindent
{\bf {APPENDIX: MAGNETIC SUSCEPTIBILITY AND DOUBLY OCCUPANCY OF THE METALLIC
STATE CLOSE TO THE MOTT-HUBBARD LOCALIZATION} }
\end{center}

We start from the expression (\ref{F(T)}) for the free-energy functional for
a constant DOS and in the half-band case:

\begin{equation}
f=\frac{F_M}{NW}=-\left( 4-m^2\right) \Phi /8+2ud^2-hm-\frac 18jm^2-\frac{%
2\pi ^2}{3\Phi }\tau ^2\text{ }+k,  \eqnum{A.1}
\end{equation}
where $\Phi $ is defined by eq. (\ref{fifi}).

Minimizing above expression with respect $m$ we obtain:

\begin{equation}
0=\frac{\partial f}{\partial m}=\frac{d^2m}{\sqrt{4\left( 1-2d^2\right)
^2-m^2}}-h-\frac 14jm+\frac{2\pi ^2}{3\Phi ^2}\Phi _m\tau ^2  \eqnum{A.2}
\end{equation}
where $\Phi _m=\partial \Phi /\partial m.$ Differentiating the above eqation
with respect $h$ and taking $h=0$, for which $m=0$, we have

\begin{equation}
\frac{d^2}{1-2d^2}m_0^{^{\prime }}-2-\frac 12jm_0^{^{\prime }}-\frac{8\pi
^2\tau ^2}3\left. \Phi _m\Phi _h\Phi ^{-3}\right| _0+\frac{4\pi ^2\tau ^2}3%
\left. \Phi _{mh}\Phi ^{-2}\right| _0=0,  \eqnum{A.3}
\end{equation}
where $m_0^{^{\prime }}=\left( dm/dh\right) _{h=0.}$ The derivative $\Phi $
for $m=h=0$ is:

\begin{eqnarray}
\left. \frac{\partial \Phi }{\partial m}\right| _0 &=&\left\{ \frac{32d^2}{%
\left( 4-m^2\right) ^2}\left( 1-2d^2+\frac 12\sqrt{4\left( 1-2d^2\right)
^2-m^2}\right) m\right. +  \nonumber \\
&&\left. -\frac{8d^2}{\left( 4-m^2\right) \sqrt{4\left( 1-2d^2\right) ^2-m^2}%
}m\right\} _{h=0}  \eqnum{A.4}
\end{eqnarray}

We se that $\Phi _{m=0}=0.$ Also, $\left( d^2\right) _0^{^{\prime }}=0;$
this is because both $\Phi $ and $d^2$ depends only on even powers of either 
$m$ or $h$. In that situation:

\begin{eqnarray}
\left. \frac{\partial ^2\Phi }{\partial m\partial h}\right| _0
&=&\allowbreak 2d^2\left( 1-2d^2+\sqrt{1-4d^2+4d^4}\right) m_0^{\prime
}-\allowbreak \frac{d^2}{\sqrt{1-4d^2+4d^4}}m_0^{^{\prime }}=  \nonumber \\
&=&\frac{4d^2\left( 1-2d^2\right) ^2-d^2}{1-2d^2}m_0^{^{\prime }}=\frac{%
d^2\left( 4d^2-3\right) \left( 4d^2-1\right) }{1-2d^2}m_0^{^{\prime }} 
\eqnum{A.5}
\end{eqnarray}

So, we obtain the magnetic susceptibility $\chi \equiv dm/dh$ in the form:

\begin{equation}
m_0^{^{\prime }}=2\left( \frac{d^2}{1-2d^2}-\frac 12j+\frac{\pi ^2\left(
4d^2-3\right) \left( 4d^2-1\right) }{48d^2\left( 1-2d^2\right) ^3}\tau
^2\right) ^{-1}.  \eqnum{A.6}
\end{equation}

We use also the low-temperature expansion of $d^2$ derived earlier:

\begin{equation}
d^2=\frac 14\left( 1-\frac u2\right) -\frac{\tau ^2\pi ^2u}{6\left( 1-\frac 1%
4u^2\right) ^2}  \eqnum{A.7}
\end{equation}

Expanding $m_0^{^{\prime }}$ to the first order in $\tau ^2$ we arrive at
the expression for $\chi $ in physical units:

\begin{equation}
\chi =\chi _0\left( \frac 1{\frac{2-u}{2+u}-j}+\frac{16}3\pi ^2u^2\frac{%
u^2+4u-4}{\left( 4-u^2\right) ^2\left( -2+u+2j+ju\right) ^2}\tau ^2\right) 
\eqnum{A.8}
\end{equation}
where $\chi _0$ is the susceptibility for noninteracting particles.

We see that for $u>-2+2\sqrt{2}\approx 0.83$ the susceptibility is always
rising with temperature. Note that for DOS smoothly varying around the Fermi
energy this increase is solely due to the correlations and corresponds to
the approaching the localization boundary depicted in Fig.7. In the
nondegenerate case and for $T=0,$ the corresponding Brinkman-Rice formula
reads:

\begin{equation}
\chi =\chi _0\frac{2+u}{2-u}  \eqnum{A.9}
\end{equation}

Now we derive $d^2$ in small magnetic field and temperature. Because of time
reversal symmetry $d^2$ depends only on even power of $h.$ We determine
nonvanishing term $\sim h^2$ at $T=0.$ Substituting into eq. (\ref{dEd})
first the expression for $m$ in the first order of $h$ given by (\ref{m2hj})
and subsequently differentiating twice, thus obtained formula we see that
doubly-occupancy probability diminishes with growing $T$ and/or $h$
according to:

\begin{equation}
d^2=\frac 14\left( 1-\frac u2\right) -\frac{\pi ^2u}{6\left( 1-\frac{u^2}4%
\right) ^2}\tau ^2-\frac 4{\left( -2+u+2j+ju\right) ^2}h^2  \eqnum{A.10}
\end{equation}

The decrease of $\eta $ with $T$ and $h$ signals a precursory localization
effects discussed in subsection {\bf II.E. }

We now generalize the results to the case with arbitrary DOS. We start from
the expression (\ref{F/2N}) for the free energy with the condition

\begin{equation}
1=\sum_\sigma \int_{-W/2}^{W/2}\rho \left( \epsilon \right) n_\sigma \left(
\epsilon \right) d\epsilon ,  \eqnum{A.11}
\end{equation}
for the chemical potential; $n_\sigma \left( \epsilon \right) $ is given by (%
\ref{nk}). Defining $H\equiv h+\frac 14Jm-\beta ,$ we obtain the following
expression of $m$ to the first order:

\begin{equation}
m=2\int_{-W/2}^{W/2}\rho \left( \epsilon \right) \sum_\sigma \sigma \left( 
\frac 1{1+e^{\frac{\phi \epsilon -\mu }{k_BT}}}+\frac 1{\left( 1+e^{\frac{%
\phi \epsilon -\mu }{k_BT}}\right) ^2}e^{\frac{\phi \epsilon -\mu }{k_BT}}%
\frac \sigma {k_BT}H\right) d\epsilon  \eqnum{A.12}
\end{equation}
or, explicitly:

\begin{equation}
m=2(h+\frac 14Jm-\beta )K,  \eqnum{A.13}
\end{equation}
with

\begin{equation}
K\equiv \int_{-W/2}^{W/2}\frac 2{\left( 1+e^{\frac{\phi \epsilon -\mu }{k_BT}%
}\right) ^2}\frac{e^{\frac{\phi \epsilon -\mu }{k_BT}}}{k_BT}\rho \left(
\epsilon \right) d\epsilon \text{ .}  \eqnum{A.14}
\end{equation}

Differentiating the above equation with respect $h$ and taking $h=0$ we have:

\begin{equation}
m_0^{^{\prime }}=2\left( 1+\frac 14Jm_0^{^{\prime }}-\beta _{h=0}^{^{\prime
}}\right) K\left( h=0\right)  \eqnum{A.15}
\end{equation}

Differentiating the free energy with respect to $m$ we obtain the relation:

\begin{equation}
\frac 14Jm-\beta +2\sum_\sigma \int_{-W/2}^{W/2}\rho \left( \epsilon \right)
\left( \Phi _m\epsilon -\sigma \frac 14J\right) n_\sigma \left( \epsilon
\right) =0,  \eqnum{A.16}
\end{equation}
and hence:

\begin{equation}
\beta =2\Phi _m\sum_\sigma \int_{-W/2}^{W/2}\epsilon \rho \left( \epsilon
\right) n_\sigma \left( \epsilon \right) ,  \eqnum{A.17}
\end{equation}
which, when differentiated with respect to $h,$ leads to:

\begin{equation}
\beta _{h=0}^{^{\prime }}=4\Phi _{mh}|_0\int_{-W/2}^{W/2}\epsilon \rho
\left( \epsilon \right) n_0\left( \epsilon \right) ,  \eqnum{A.18}
\end{equation}
where $\Phi _{mh}|_0$ is given by (A.5). From (A.15) we have:

\begin{equation}
\left. \frac{dm}{dh}\right| _{h=0}=2\frac K{-\frac 12KJ+4K\bar{\epsilon}%
\left( T\right) \frac{d^2\left( 4d^2-3\right) \left( 4d^2-1\right) }{1-2d^2}%
+1}  \eqnum{A.19}
\end{equation}
where all quantities on right hand side are taken for $h=0,$ and $\bar{%
\epsilon}\left( T\right) $ is average band energy per site and orbital.

Next, we expand $\bar{\epsilon}\left( T\right) ,$ $d^2$ and $K$ in the
powers of $T.$ From \cite{20} we have that:

\begin{equation}
d^2=\frac 14\left( 1-I\right) -\frac{2\pi ^2}3\frac{I\rho }{U_0}\left( \frac{%
k_BT}{\Phi _0}\right) ^2,  \eqnum{A.22}
\end{equation}
\begin{equation}
\Phi =\Phi _0-\frac{16\pi ^2}3\frac{I^2\rho }{U_0}\left( \frac{k_BT}{\Phi _0}%
\right) ^2,  \eqnum{A.23}
\end{equation}
and 
\begin{equation}
\bar{\epsilon}\left( T\right) =-\frac{U_0}8+\frac 13\pi ^2\rho \left( \frac{%
k_BT}{\Phi _0}\right) ^2.  \eqnum{A.24}
\end{equation}
where $I\equiv U/U_0,$ and $U_0\equiv 8\left| \bar{\epsilon}\right| .$
Defining $\mu ^{*}=\mu /\Phi ,$ and $T^{*}=T/\Phi ,$ we can write

\begin{eqnarray}
K &=&\int_{-W/2}^{W/2}\frac 2{\left( 1+e^{\frac{\epsilon -\mu ^{*}}{k_BT^{*}}%
}\right) ^2}\frac{e^{\frac{\epsilon -\mu ^{*}}{k_BT^{*}}}}{k_BT}\rho \left(
\epsilon \right) d\epsilon =  \nonumber \\
&=&-\frac 2\Phi \int_{-W/2}^{W/2}f^{^{\prime }}\left( \frac{\epsilon -\mu
^{*}}{k_BT^{*}}\right) \rho \left( \epsilon \right) d\epsilon =  \nonumber \\
&=&-\frac 2\Phi \int_{\frac{-W/2-\mu ^{*}}{k_BT^{*}}}^{\frac{W/2-\mu ^{*}}{%
k_BT^{*}}}f^{^{\prime }}\left( x\right) \rho \left( xk_BT^{*}+\mu
^{*}\right) dx=  \nonumber \\
&=&-\frac 2\Phi \sum_n\left( k_BT^{*}\right) ^{2n}\frac{\rho ^{\left(
2n\right) }\left( \mu ^{*}\right) }{\left( 2n\right) !}\int_{-\infty
}^{+\infty }x^{2n}f^{^{\prime }}\left( x\right) dx  \eqnum{A.25}
\end{eqnarray}
where $f^{^{\prime }}\left( x\right) $ is the derivative of the Fermi-Dirac
function. Noting that

\begin{eqnarray}
\int_{-\infty }^{+\infty }\frac{df\left( x\right) }{dx}dx &=&-1,  \nonumber
\\
\int_{-\infty }^{+\infty }x^2\frac{df\left( x\right) }{dx}dx &=&-\frac 13\pi
^2,  \eqnum{A.26}
\end{eqnarray}
we arrive at the expression

\begin{equation}
K=\frac 2\Phi \left[ \rho \left( \mu ^{*}\right) +\frac 16\pi ^2\rho
^{^{\prime \prime }}\left( \mu ^{*}\right) \left( k_BT^{*}\right) ^2\right] ,
\eqnum{A.30}
\end{equation}

From \cite{20} we also have:

\begin{equation}
\mu ^{*}\left( T\right) =\epsilon _F-\frac{\pi ^2}6\left( k_BT^{*}\right) ^2%
\frac{\rho ^{^{\prime }}}\rho ,  \eqnum{A.31}
\end{equation}
and thus

\begin{eqnarray}
K &=&\frac{2\rho }\Phi \left[ 1-\frac{\pi ^2}6\left( \frac{k_BT}\Phi \right)
^2\left( \left( \frac{\rho ^{\left( 1\right) }}\rho \right) ^2-\frac{\rho
^{\left( 2\right) }}\rho \right) \right] =  \nonumber \\
&=&2\rho \left( \frac 1{\Phi _0}+\frac{16\pi ^2}{3\Phi _0^2}\frac{I^2}{U_0}%
\rho \left( k_BT^{*}\right) ^2\right) -2\rho \frac{\pi ^2}{6\Phi _0}\left(
k_BT^{*}\right) ^2r,  \eqnum{A.32}
\end{eqnarray}
with $r=\left( \frac{\rho ^{\left( 1\right) }}\rho \right) ^2-\frac{\rho
^{\left( 2\right) }}\rho .$ We substitute (A.22), (A.24) and (A.32) into
(A.19). In result, the susceptibility $\chi =\frac 1{4\rho }\chi _0\left(
dm/dh\right) _{h=0}$ is of the form:

\begin{equation}
\chi =\frac{\chi _0}S\left[ 1-a\frac{\pi ^2\left( k_BT\right) ^2}{6\Phi _0^2S%
}\right] ,  \eqnum{A.33}
\end{equation}
where

\begin{equation}
S=\Phi _0\left( 1-U\rho \frac{1+I/2}{\left( 1+I\right) ^2}\right) -J\rho , 
\eqnum{A.34}
\end{equation}
and

\begin{equation}
a=8I^2\left( I^2+2I+3\right) \left( 1+I\right) ^{-2}\rho ^2-\frac{32I^2\rho 
}{Uc}+r\Phi _0.  \eqnum{A.35}
\end{equation}

Neglecting higher-order contribution in $I,$ the result reduces to the usual
Stoner form $\chi \left( T=0\right) =\chi _0\left[ 1-\rho \left( U+J\right)
\right] ^{-1}.$ Formula (A.33) generalizes the result of Brinkman and Rice 
\cite{8} obtained for a nondegenerate case at $T=0,$ and its generalization 
\cite{20} to $T>0.$ The Hund's rule coupling enhances the susceptibility and
can be the source of a ferromagnetic instability.

With the help of the expression for $\Phi $ one can determine the first
nontrivial contribution to the specific heat in the applied field. Namely,
one has $C_v=\left( \gamma _0/\Phi _0\right) T+bH_a^2T$, where $b$ is a
constant divergent at the localization point. This increase is connected
with the narrowing of the band with the growing field ($\Phi \sim d^2$).

\begin{center}
\newpage \noindent{\large {\bf Figure Captions}}
\end{center}

\vspace{1cm} \noindent
{\bf Fig.1 }Field dependences of double occupancy $d^2$ per orbital (a) and
magnetization $m$ per atom (b), both for $T=0$ and $n=2.$\vspace{0.5cm}

\noindent
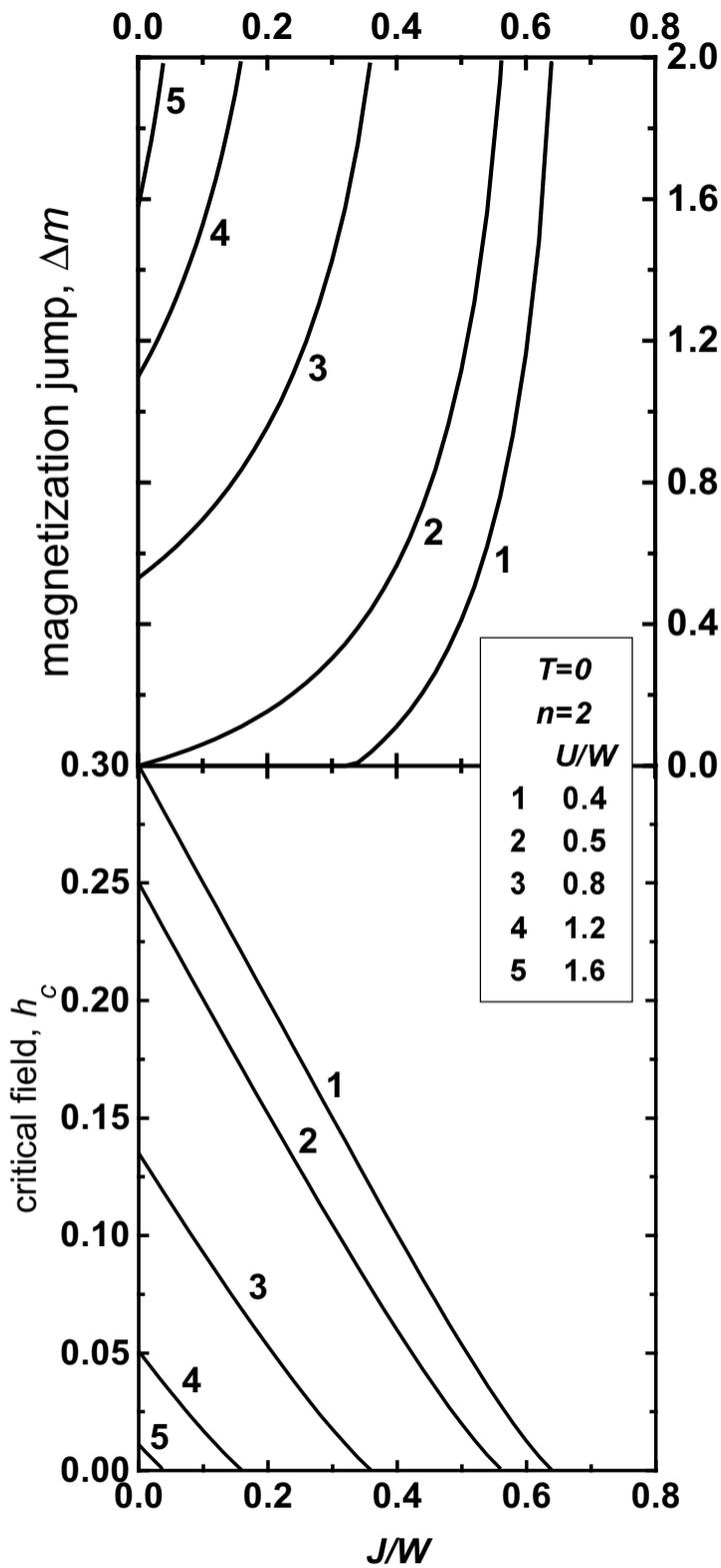
{\bf Fig.2 }Jump of the magnetization $\Delta m$ per atom and the critical
magnetic field $h_c$ both as a function of exchange interaction $J/W$ and
calculated at the metal-insulator transition for $T=0$ and $n=2.$\vspace{%
0.5cm}

\noindent
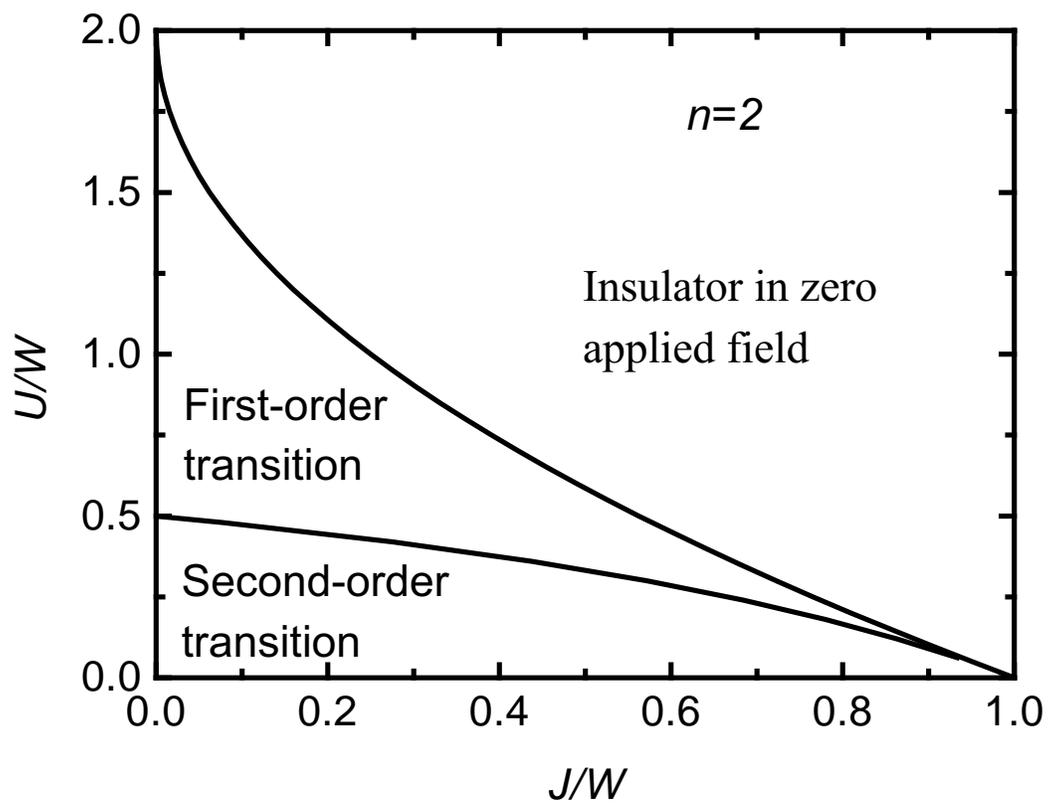
{\bf Fig.3 }Regimes of $U/W$ and $J/W$ for the half-filled band
configuration comprising both first- and second-order transitions regimes in
an applied magnetic field.\vspace{0.5cm}

\noindent
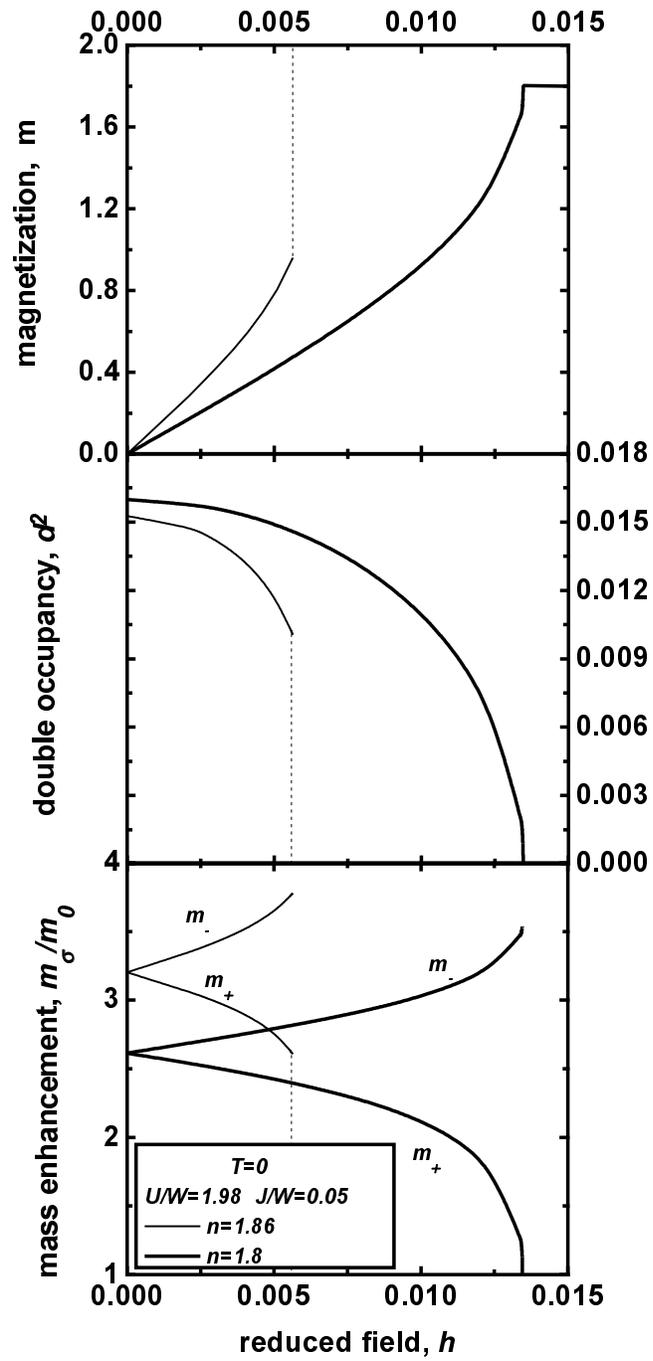
{\bf Fig.4 }Field dependences of magnetization $m$ per atom, double
occupancy $d^2$ per orbital, and the spin-dependent mass enhancement $%
m_\sigma /m_0$, all for $n<2$ (from the top to the bottom, respectively)$.$%
\vspace{0.5cm}

\noindent
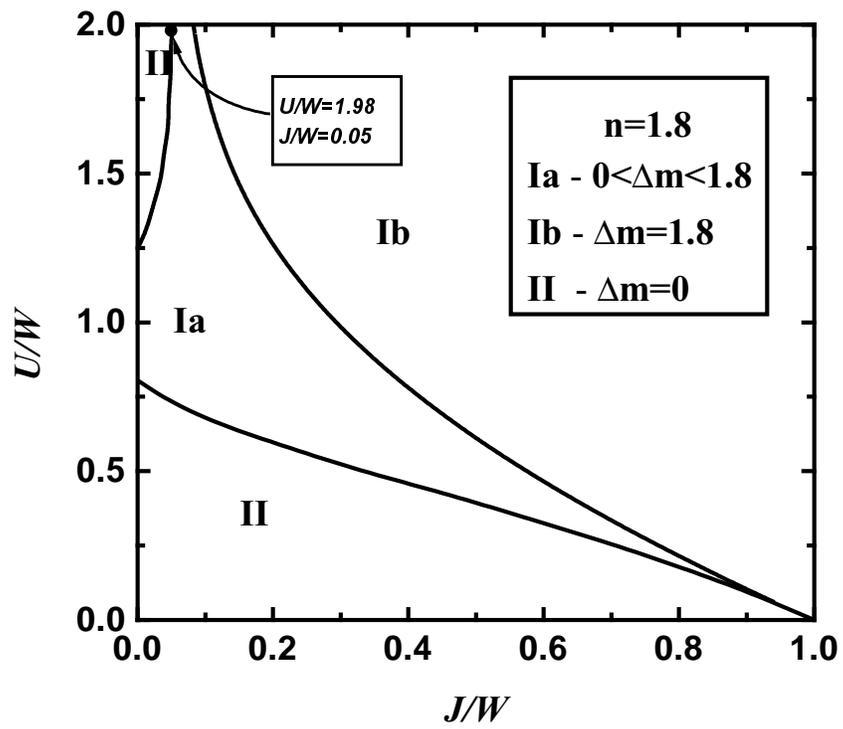
{\bf Fig.5 }Regimes of $U/W$ and $J/W$ of first- and second-order
transitions in applied magnetic field for $T=0.$\vspace{0.5cm}

\noindent
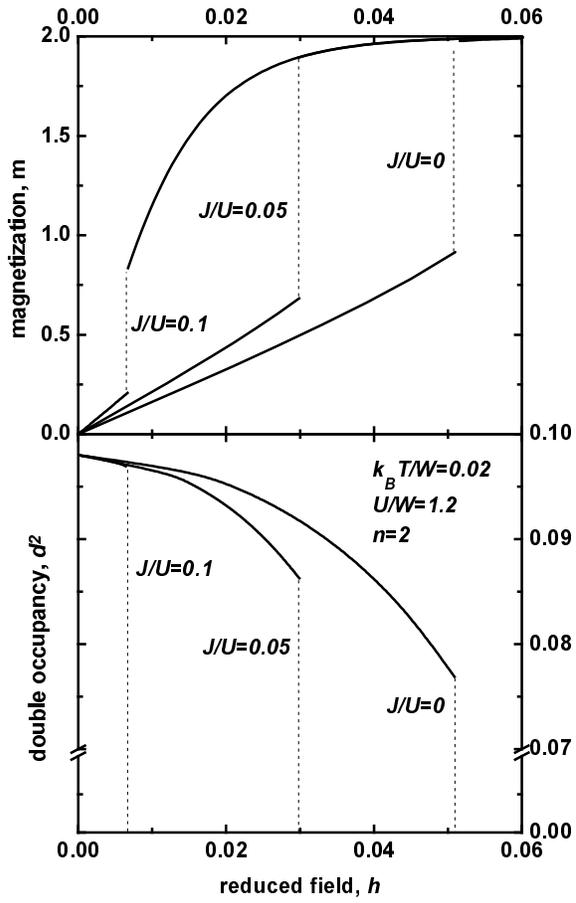
{\bf Fig.6 }Field dependences of the magnetization and double occupancy for
different values of $J/U.$\vspace{0.5cm}

\noindent
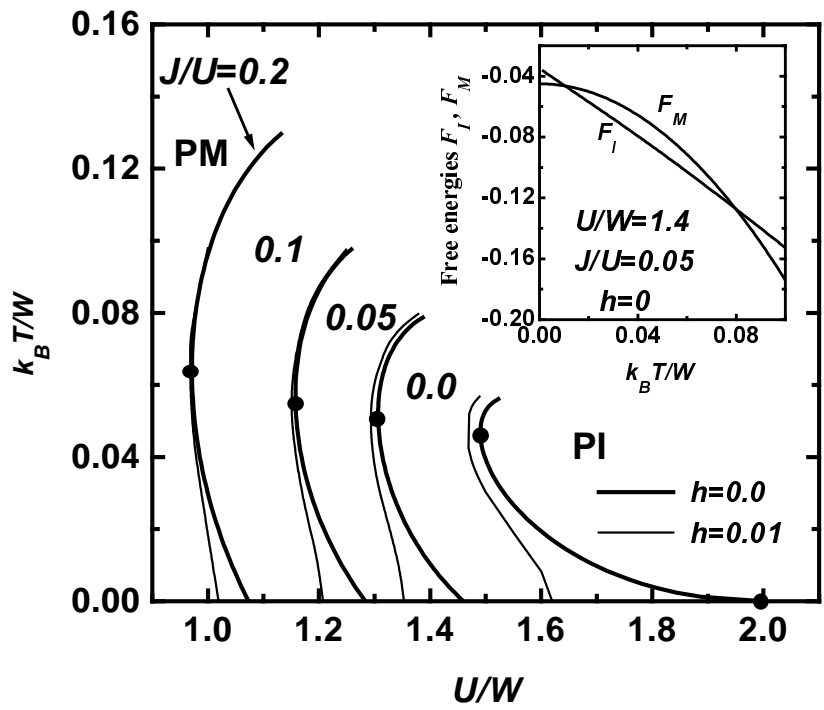
{\bf Fig.7 }The paramagnetic metal (PM) - paramagnetic insulator (PI) phase
boundaries as a function of $U/W$ for $n=2$ (thick lines), and their shift
in the applied magnetic field (fine lines). The inset displays the free
energies for PM and PI states as a function of temperature.\vspace{0.5cm}

\noindent
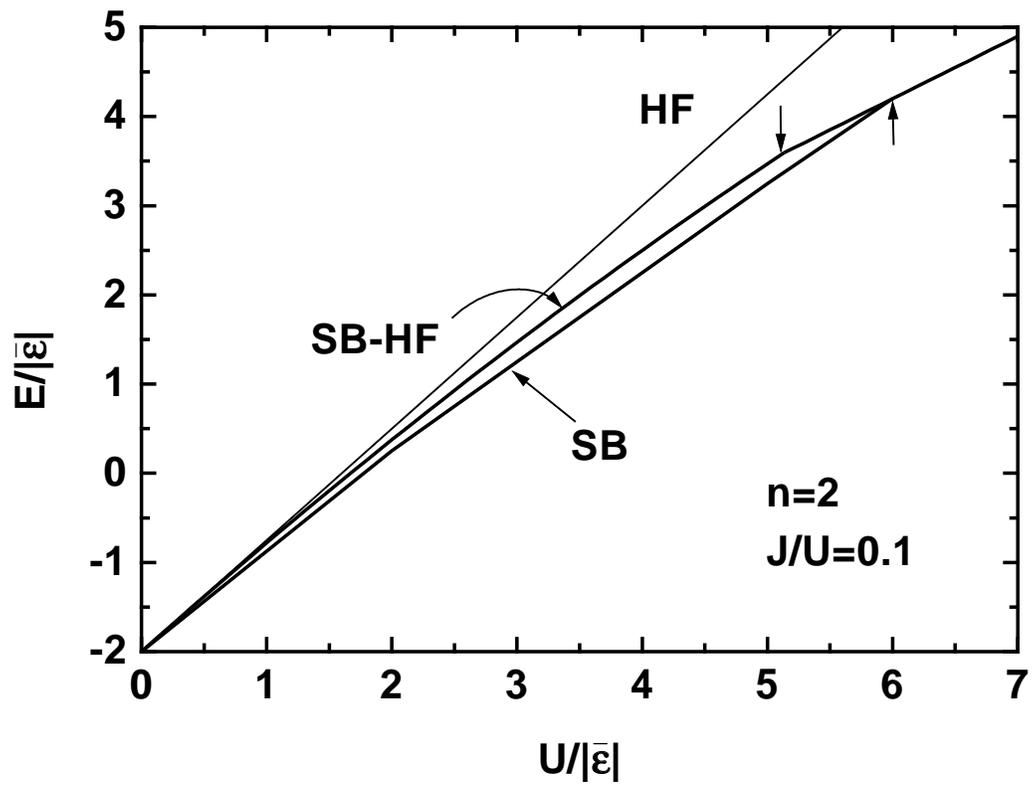
{\bf Fig.8 }Comparison ground state energies for $T=h=0,$ $n=2$ and $J/U=0.1$
obtained from the methods: Hartree-Fock (HF), slave boson(SB) and slave
boson combined with Hartree-Fock (SB-HF). The arrows indicate the position
of MIT in the two last approximation schemes.\vspace{0.5cm}

\noindent
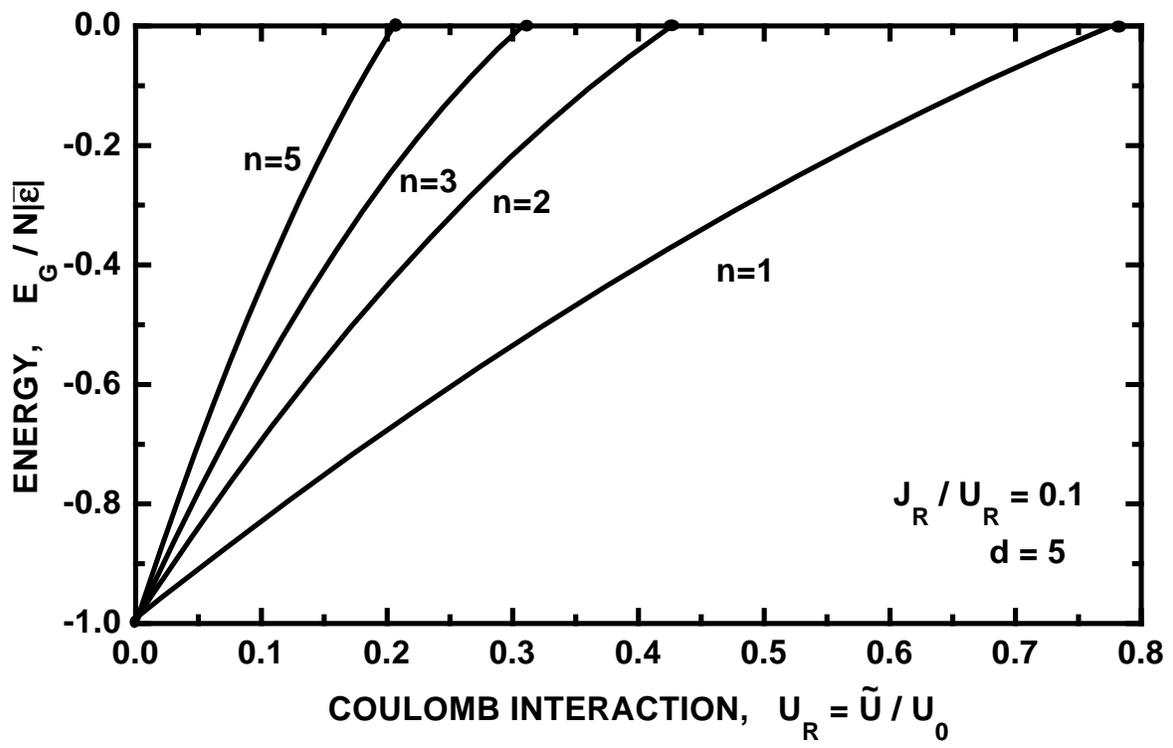
{\bf Fig.9 }The ground energy per site (in units of $\left| \bar{\epsilon}%
\right| $) as a function of the Coulomb interaction $U_R=\tilde{U}/U_0\equiv 
\tilde{U}/8\left| \bar{\epsilon}\right| .$ A transition to the Mott-Hubbard
insulating state takes place for each (integer) band filling $n$ for the
value of $U_R$ at which $E_G=0.$\vspace{0.5cm}

\noindent
{\bf Fig.10 a) }The relative magnitude of the charge fluctuations $\lambda
_R=\left( \lambda -\lambda _0\right) /\left( \lambda _\infty -\lambda
_0\right) $ versus $U_R.$ Note a discontinuous nature of the transition to
the atomic configuration $\lambda =\lambda _\infty $ at a critical value of $%
U_R$ specified for each $n>1.$

\quad \quad {\bf b)} The relative magnitude of the local moment $m_R=\left(
m-m_0\right) /\left( m_\infty -m_0\right) ,$ as a function of $U_R.$ The
discontinuous jump in $m$ reflects the same behavior as of $\lambda _R.$%
\vspace{0.5cm}

\noindent
{\bf Fig.11 a)} The band-narrowing parts $\Lambda \left( \lambda \right) $
and $G\left( m\right) $ as the magnitude of the Coulomb interaction. The
inset compares the two narrowing factors, both diminishing with increasing $%
U_R.$

\quad \quad {\bf b) }The charge- and spin-fluctuation parts of the band
narrowing as a function of the Coulomb interaction, for the value of $U_R$
below the Mott transition. The inset displays the different trend as a
function $J/\tilde{U}$ ratio.

{\bf \ }

\end{document}